\documentclass[longauth]{aa} 

\usepackage{soul}
\usepackage{color, colortbl}
\usepackage{tikz}
\usetikzlibrary{patterns}
\usepackage{graphicx}
\usepackage{subcaption}

\usepackage{txfonts}

\usepackage{hyperref}
\begin{document}

   \title{An impressionist view of V Hydrae}

   \subtitle{When MATISSE paints Asymmetric Giant Blobs\thanks{Based on observations made at the Paranal Observatory under ESO Programmes ID 108.22E9, 096.D-0568, and 079.D-0140(A).}} 

\author{L. Planquart
          \inst{1,} \inst{2} \fnmsep\thanks{
Research fellow, 
FNRS, Belgium.}
          \and C. Paladini \inst{2}
          \and A. Jorissen \inst{1}
          \and A. Escorza \inst{2, }\inst{3,} \inst{4}    
          \and E. Pantin \inst{5}
          \and J. Drevon \inst{2, }\inst{6}
          \and B. Aringer\inst{7}
          \and F. Baron \inst{8}
          \and A. Chiavassa \inst{6}
          \and P. Cruzal\`ebes \inst{6}
          \and W. Danchi \inst{9}
          \and E. De Beck \inst{10} 
          \and M. A. T. Groenewegen \inst{11}
          \and S. {H{\"o}fner} \inst{12}
          \and J. Hron \inst{7}
          \and T. Khouri\inst{10}
          \and B. Lopez \inst{6}
          \and F. Lykou \inst{13}
          \and M. Montarges \inst{14}
          \and N. Nardetto \inst{6}
          \and K. Ohnaka \inst{15}
          \and H. Olofsson \inst{10}
          \and G. Rau \inst{9}
          \and A. Rosales-Guzm\'an  \inst{16}
          \and J. Sanchez-Bermudez \inst{16}
          \and P. Scicluna \inst{2}
          \and L. Siess \inst{1}
          \and F. Th\'evenin \inst{6}
          \and S. Van Eck \inst{1}
          \and W.H.T. Vlemmings \inst{10}
          \and G. Weigelt \inst{17}
          \and M. Wittkowski \inst{18} 
          }

   \institute{Institut d’Astronomie et d’Astrophysique, Université Libre de Bruxelles, CP 226, Boulevard du Triomphe, 1050 Brussels,
Belgium \\ \email{lea.planquart@ulb.be}
         \and
             European Southern Observatory, Alonso de Córdova, 3107 Vitacura, Santiago, Chile
        \and
        Instituto de Astrofísica de Canarias, C. Vía Láctea, 38205 La Laguna, Tenerife, Spain
\and
Universidad de La Laguna, Departamento de Astrofísica, Av. Astrofísico Francisco Sánchez, 38206 La Laguna, Tenerife, Spain
\and
        Université Paris-Saclay, Université Paris Cité CEA, CNRS, AIM, 91191, Gif-sur-Yvette, France
\and
Université Côte d’Azur, Observatoire de la Côte d’Azur, CNRS, Laboratoire Lagrange, France
\and
Department of Astrophysics, University of Vienna, T\"{u}rkenschanzstraße 17, A-1180 Wien, Austria
\and
Center for High Angular Resolution Astronomy and Department of Physics and Astronomy, Georgia State University, P.O. Box 5060, Atlanta, GA 30302-5060, USA
\and
NASA Goddard Space Flight Center, United States
\and 
Department of Space, Earth and Environment, Chalmers University of Technology, Onsala Space Observatory, 43992 Onsala, Sweden
\and
Royal Observatory of Belgium, Ringlaan 3, 1180 Brussels, Belgium 
\and
Theoretical Astrophysics, Department of Physics and Astronomy, Uppsala University, Box 516, 751 20 Uppsala, Sweden 
\and Konkoly Observatory, Research Centre for Astronomy and Earth Sciences, Hungarian Research Network (HUN-REN), Konkoly Thege Miklós út 15-17., H-1121 Budapest, Hungary
\and LESIA, Observatoire de Paris, Université PSL, CNRS, Sorbonne Université, Université Paris Cité, 5 place Jules Janssen, 92195 Meudon, France
\and Instituto de Astrofísica, Universidad Andrés Bello, Fernández Concha 700, Las Condes, Santiago, Chile
\and Instituto de Astronomía, Universidad Nacional Autónoma de México, Apdo. Postal 70264, Ciudad de México, 04510, México
\and Max Planck Institute for Radio Astronomy, Auf dem Hügel 69,
53121 Bonn, Germany
\and European Southern Observatory (ESO), Karl-Schwarzschild-Str. 2, D-85748 Garching bei München, Germany
             }

   \date{Received 10 October 2023 / Accepted 1 May 2024}
  \abstract
   {Asymptotic Giant Branch (AGB) stars enrich the interstellar medium through their mass loss. The mechanism(s) shaping the circumstellar environment of mass-losing stars is not yet clearly understood.}
   {Our purpose is to study the effect of binary companions located within the first 10 stellar radii from the primary AGB star. In this work, we target the mass-losing carbon star V~Hydrae (V~Hya), looking for signatures of its companion in the dust forming region of the atmosphere.}
   {The star was observed in the $L$- and $N$-bands with the VLTI/MATISSE instrument at low spectral resolution. We reconstructed images of V~Hya's photosphere and surroundings using the two bands and compared our interferometric observables with VLTI/MIDI and VISIR archival data. To constrain the dust properties, we used the 1D-radiative transfer code DUSTY to model the spectral energy distribution.} 
   {The star is dominated by dust emission in the $L$- and $N$- bands. The MATISSE reconstructed images show asymmetric and elongated structures in both infrared bands. In the $L$-band, we detected an elongated shape of approximately 15~mas, likely to be of photospheric origin. In the $N$-band, we found a 20~mas extension North-East from the star, and perpendicular to the $L$-band elongated axis.
   The position angle and the size of the $N$-band extension match the prediction of the companion position at MATISSE epoch. 
   By comparing MATISSE $N$-band with MIDI data, we deduce that the elongation axis in the $N$-band has rotated since the previous interferometric measurements 13 years ago, supporting the idea that the particle enhancement is related to the dusty clump moving along with the companion. The VISIR image confirms the presence of a large-scale dusty circumstellar envelope surrounding V~Hya.}
  {The MATISSE images unveil the presence of a dust enhancement at the companion position, opening new doors for further analysis on the binary interaction with an AGB component.}

   \keywords{stars: AGB and post-AGB -- stars: carbon -- technique: interferometry -- stars:individual: V~Hya -- stars: mass-loss
               }

   \maketitle
%
\section{Introduction}
\label{sect:intro}
During the Asymptotic Giant Branch (AGB) phase, low- to intermediate-mass stars eject most of their mass through stellar winds ($\dot{M} >10^{-8}$~M$_{\odot}$~yr$^{-1}$, \citealt{Hofner_Olofsson_2018A&ARv..26....1H}), contributing to the dust and chemical enrichment of the Galaxy. 
When they begin their transition from the AGB to the pre-planetary nebula (pPN) or post-AGB phase, a variety of complex geometries arises. 
The emergence of asymmetric outflows in these final stages of stellar evolution (multipolar jets, spirals, or discs) is believed to be driven by the interaction of the AGB environment with a binary companion (\citealt{PN_claire_ref_2017NatAs...1E.117J}, \citealt{2020Sci...369.1497D}).
However, there are two main limitations in trying to detect binary companions of evolved stars: on one hand, the primary is close and much brighter (a few hundred to a thousand times) than the secondary, limiting direct-imaging detection of the companion. On the other hand, the AGB star often exhibits strong variability related to pulsation, making indirect detection of the companion through radial-velocity difficult as well. Only a few AGB binaries have been detected so far and very few of them have been studied in depth (e.g. \citealt{Mira_ALMA_2014A&A...570L..14R}, \citealt{Kervella_2016A&A...596A..92K}, \citealt{WAql_2020A&A...633A..13D}). Direct observations of such systems are nevertheless needed to improve our understanding of the shaping mechanism of the stellar wind. 
The BIN-AGB ESO Large Program (nickname for BINary AGBs, PI Paladini, prog ID: 108.22E9) aims at imaging the first 10 stellar radii of 10 different AGB stars with different chemistry and variability class. Many of the targets, including the one here presented, are suspected binaries, and the main aim is to observe how the presence of a companion affects the onset and the morphology of the stellar wind.

V Hydrae (\object{V~Hya}) is a carbon star which exhibits peculiar properties for an AGB star: a high mass-loss rate ($10^{-5}$~M$_{\odot}$~yr$^{-1}$, \citealt{Knapp_1997A&A...326..318K}), a mean spectral broadening of 13.5~$\rm km\,s^{-1}$ with a periodic variation of 9~$\rm km\,s^{-1}$ \citep{Barnbaum_1995ApJ...450..862B}, and a dense equatorial disc combined with a bipolar outflow mapped by the radio emission of CO J~=~2–1 and J~=~3–2 (from a 15\arcsec~spatial mesh). The latter suggests that the star may be at the transition from the AGB phase to a planetary nebula, and its environment is shaped by a companion  \citep{Tsuji,Kahane_1996A&A...314..871K, referee_suggestion_Hirano_CO_radio_wind_2004ApJ...616L..43H}.
Distance estimates to V~Hya range from 340~pc to 550~pc (\citealt{Barnbaum_1995ApJ...450..862B}, \citealt{Bergeat_1998A&A...332L..53B}) and the measured parallax of $2.31\pm0.11$~mas corresponds to a distance of $434\pm21$~pc \citep{GDR3_2022arXiv220800211G}, which is adopted in the current paper.\footnote{\cite{Gaia_AGB_priors_2022A&A...667A..74A} reports that the uncertainties on Gaia distances for red stars are underestimated by up to a factor 5. Hence, caution should be exercised when converting parallaxes into distances by a simple inversion, since the parallax resulting from this inversion becomes largely biased when the parallax fractional error is larger than 0.2 \citep{Bailer-Jones2015}.}




Photometric monitoring in the visual band shows a nearly-sinusoidal variation with a period of 530~$\pm$~30 days, typical of Mira-type stars, as well as a periodic dimming event every 17~$\pm$~1 yr \citep{KNAPP_1999A&A...351...97K}. The latter authors interpreted the dimming event as being caused by a thick cloud orbiting the star and connected to the binary companion. Hints of the presence of a companion were also revealed by the detection of forbidden emission lines in the blue part of the optical spectrum \citep{Lloyd_evans_1991MNRAS.248..479L} and by an excess of UV flux that \cite{Sahai_2008ApJ...689.1274S} explained as arising from a hot accretion disc surrounding the companion. Such an accretion disc resulting from mass transfer in a binary system was also suggested as the origin of the varying high-velocity (up to 120~$\rm km\,s^{-1}$) jet seen in the 4.6~$\mu$m CO vibration-rotation band  \citep{referee_suggestion_Sahai_CO_outflow_paper_1988A&A...201L...9S, referee_suggestion_Sahai_CO_outflow_2_2009ApJ...699.1015S} and in optical atomic lines \citep{Lloyd_evans_1991MNRAS.248..479L,self_citation}. From a long-term radial-velocity monitoring combined with proper-motion acceleration, \cite{self_citation} proposed an orbital solution with an orbital period of 17.5~yr, a low eccentricity, and an inclination angle of 40$^\circ$ between the orbital plane and the plane of sky, leading to a separation of about 11~au. Given the orbital model, the periodic dimming event coincides with the superior conjunction and is explained by a dusty clump moving with the companion, whereas the blueshifted absorption lines are explained by the presence of a conical high-velocity jet launched at the companion position.

Previous imaging studies already demonstrated the complex circumstellar environment around the star at different scales, however, none of them directly identified the companion. Starting from the large spatial scales at the interface with the interstellar medium: ALMA images at an angular resolution of 0.5\arcsec\ (220~au adopting a distance of 434~pc) by \cite{SahaiALMA} reveal the presence of concentric rings and bipolar arcs surrounding the star in the sub-millimeter range. HST images taken at six different epochs show that the system undergoes periodic ejections of "bullets", in the form of high-velocity collimated material seen  in the emission line of [\ion{S}{II}]  \citep{Sahai_HST}.

With the Gemini Planet Imager instrument (GPI), \cite{Sahai_2019IAUS..343..495S} imaged the polarized (scattered) light of an extended central dusty disc (radius $\sim$ 0.5\arcsec) in the near-infrared ($Y$-band). 
The dusty environment was imaged in the thermal infrared ($N$-band) with the mid-infrared TIMMI camera by \cite{Lagadec}, revealing a hot-spot offset from the star, $0.9\arcsec$ in the North-East direction. Additionally, interferometric observations with higher angular resolution in the same infrared band were also performed.  \cite{Townes2011}, using short baselines (4~m to 35~m) configuration, detected a central source dominated by an extended environment of radius about 280~mas. 
\citealt{Zhao_Geisler}, using baselines ranging up to 70~m, confirm the presence of an extended  background and report the complexity of the data at larger baselines.


In this paper, we report the first polychromatic images of the star V~Hya and its close environment (between 1 and 10 stellar radii) obtained from interferometric observations in the $L$- and $N$-bands. 
The paper is structured as follows: in Section 2 we summarize the observational setup and the data reduction procedure. In Sects. 3 and 4, we describe the visibility fitting and the image reconstruction strategy, respectively.
In Sect. 5, we fit the spectral energy distribution of V~Hya to derive the dust properties of the circumstellar environment. In Sect. 6, we present our main results, including the images of V~Hya and its complex circumstellar environment. 
In Sect. 7, we discuss the origin of the geometry of the circumstellar environment and the asymmetric structures within it. In Sect. 8, we provide concluding remarks. 

\section{Observations and data reduction}
\label{sect:observation}

\subsection{MATISSE data}
\label{subsect:matissedata}
The main dataset presented in this paper were obtained using the Multi Aperture Mid-Infrared SpectroScopic Experiment (MATISSE) instrument \citep{MATISSE}. MATISSE works simultaneously in three infrared bands -- $L$ ($3.0$ to $4.0~\mu$m), $M$ ($4.5$ to $5.0~\mu$m) and $N$ ($8$ to $13~\mu$m) -- and combines the beams of four telescopes of the Very Large Telescope (VLT, at Cerro Paranal, Chile). These telescopes are either Unit Telescopes (UTs, 8~m aperture) or the Auxiliary Telescopes (ATs, 1.8~m aperture). 
The AT telescopes can be moved on the VLT platform.
They span baselines from about 12 up to 140 meters\footnote{Since 2023, the 200~m baseline configuration is offered to the community.}. These configurations allow us to cover different regions of the ($u,v$)-plane, reaching an unprecedented angular resolution in the $L$-band ($\lambda/2B_{max}$ = 2.2~mas, at $\lambda$ = $3~\mu$m and $B_{max}$ = 140 m), and in the $N$-band (up to 7~mas at $\lambda$ = $10~\mu$m). 

The target V~Hya was observed during the first semester of 2022 as part of the Large Program 108.22E9. The complete journal of observations can be found in Table \ref{tab:log_observation}. 
The goal of the program is to image the overall geometry of the various targets, hence the $L$- and $N$-bands low spectral resolution ($R$~=~30) setup was chosen.
The $L$-band data of V~Hya are obtained using the slow detector reading speed and a detector integration time (DIT) of $0.075~$s with central wavelength 3.5~$\mu$m. At the time of the observations this detector setup was the only one optimised to cover entirely the C$_2$H$_2$+HCN absorption features in the spectrum of C-rich AGB stars \citep{2000A&A...356..253J}. However, it was limited to the wavelength range $2.85- 4.20~\mu$m and did not allow us to observe at $M$-band.
The data analysis presented here starts at 3~$\mu$m because data at shorter wavelengths are too noisy. The $N$-band data are taken with the only detector mode available for the low spectral resolution setup (high gain and DIT$ = 0.020~$s) allowing us to cover the full $N$-band wavelength range. 

The time interval between the first and the last observation night was 93 days, corresponding to a phase span of 0.17 (over the 530 days pulsation period, \citealt{KNAPP_1999A&A...351...97K}). As we demonstrate below based on the integration of the RV-curve to derive diameter variations, this variation remains below the resolution of MATISSE as long as the time span does not exceed one fifth of the period. As a consequence, the effect of the pulsation is ignored in the data analysis, and all the data are merged together.
The peak-to-peak radial-velocity amplitude measured in the optical is about 10.7~$\rm km\,s^{-1}$ over the pulsation cycle \citep{Barnbaum_1995ApJ...450..862B}. Assuming that the radial pulsation follows a sinusoidal evolution, such a velocity amplitude would correspond to a stellar radius variation of about 0.5~au over the entire cycle \citep{self_citation}. 
The radius evolution during the MATISSE observing span would correspond to $\pm$0.3~mas (taking a distance of 340~pc as a lower boundary), well below the angular resolution of the MATISSE observations.
Additionally, estimates of radius variation for AGB stars in the thermal infrared through interferometric measurements were performed by  \cite{Zhao_Geisler1} for the SRa/Mira star W Hya. 
\cite{Zhao_Geisler1} concluded in a weak pulsation dependency (5.4$\pm$1.8~mas) for the diameter variation between visual minimum and maximum. Translating this to V~Hya's distance (W Hya's distance being estimated around $98_{-18}^{+30}$~pc, \citealt{WHya_distance_2003A&A...407..213V}), an overall diameter variation of 1.8$\pm$0.6~mas would be expected, in accordance with the estimates obtained from the variation of the RV observations. 
\begin{figure}[t]
    \includegraphics[width = 0.5\textwidth]{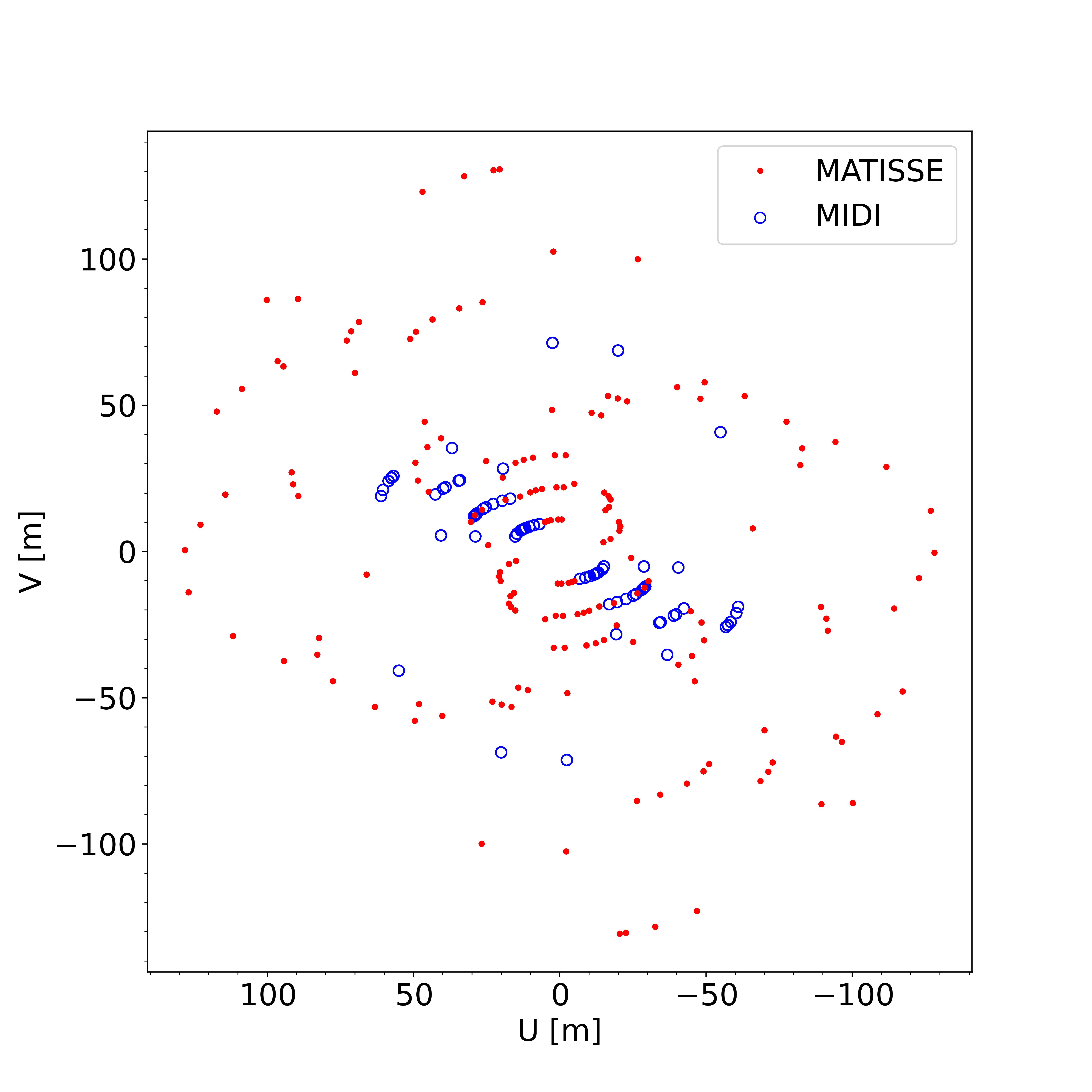}
    \caption{($u,v$)-plane coverage of MATISSE (red dots) and MIDI observation (blue open circles). U and V represent the coordinates of the sky-projected baseline vectors (expressed in meters).}
    \label{fig:UV}
\end{figure}

Each observing block of the science target is preceded and/or followed by the observation of a calibrator star. 
 Table \ref{tab:cali_info} lists the calibrators, the spectral type, their $L$- and $N$-band diameters, and their corresponding fluxes.
The raw data were processed using the python tools mat\_tools\footnote{Freely available at https://gitlab.oca.eu/MATISSE/
tools/wikis/home} provided by the MATISSE consortium. The python interface encompasses the standard \textsc{esorex} (version 3.13.5) data reduction tool for ESO/MATISSE.
The data reduction was performed following the standard steps described in \cite{2022A&A...665A..32D}.
Due to the different geometrical set-up for the $N$- and $L$-band detectors, the phase convention of the $N$-band is flipped with respect to the one of the $L$-band. As of version 1.5.2 of the DRS, this results in a sign inversion for the $N$-band closure phase and hence a 180° rotation in the final reconstructed images (see Sect. 4.8 in \citealt{AGN_2022Natur.602..403G}). The images presented in this paper have been rotated accordingly, to project them correctly onto the sky plane.

Figure \ref{fig:vis2t3_all} shows the squared visibilities and closure phases as a function of the spatial frequencies. The plot of the squared visibility shows a lack of data points above the 0.3 level in both bands, due to the presence of an extended thermal emission background over-resolved by the instrument. The closure phase plot exhibits a non-zero signal, revealing that the source is not centro-symmetric.

\subsection{VISIR data}
\label{subsect:visirdata}

To get a complementary view on V~Hya in the $N$-band, we also used archival data from the VLT mid-infrared spectro-imager VISIR  \citep{VISIR_2004Msngr.117...12L}. 
V~Hya and the point spread function calibrator (hereafter PSF) HD\,93813 were observed on 30 January 2016  (Program: 096.D-0568) in coronographic imaging mode ($\lambda_0$ = 11.3~$\mu$m, $\Delta \lambda$= 0.5~$\mu$m,  filter name = 11$\_$3$\_$4QP).


The annular groove phase mask coronagraph optimized to work in the $N$-band was used \citep{2012SPIE.8446E..8KD}. The total point source rejection of this type of vortex coronagraph is typically around
50. The observational procedure was identical to standard imaging, i.e., chopping and nodding in parallel mode to suppress the large thermal background. After chopping/nodding correction, the resulting images are
background-noise limited. A total integration time of 600 s was dedicated to V~Hya, half of which was spent with the object under the coronagraph (the rest of the time, the object is out of the coronagraph center due to the
chopping/nodding sequence). After a standard data reduction using a custom-made pipeline \citep{Pantin_2010PhDT.......249P}, the high-level datacubes were analyzed and sorted out. The worst frames (based on centering under the coronagraph and point source leaks) were rejected using an automatic selection algorithm. Several levels of rejection yield (30\%, 60\%, 90\%) were tested and the resulting images can be found in Appendix \ref{appendix:visirdata}. A visual inspection of the results after frame selection 
showed very similar structures, and the rejection yield with the best signal-to-noise ratio (90\%) was retained so that the final image was obtained by disregarding the 10\% worst frames of the datacube. This image was then corrected for PSF residuals using observations of the standard star \object{HD93813}, assumed to be a point source.
The final data were then flux-calibrated using the standard star (from the infrared calibrators list of \citealt{2003AJ....125.2645C}) where HD\,93813 is assumed to have a flux density of 27.83~Jy in the filter used.

\subsection{MIDI data}
\label{subsect:mididata}
\begin{figure}[t]
     \centering
     \includegraphics
     [width = 0.5\textwidth]{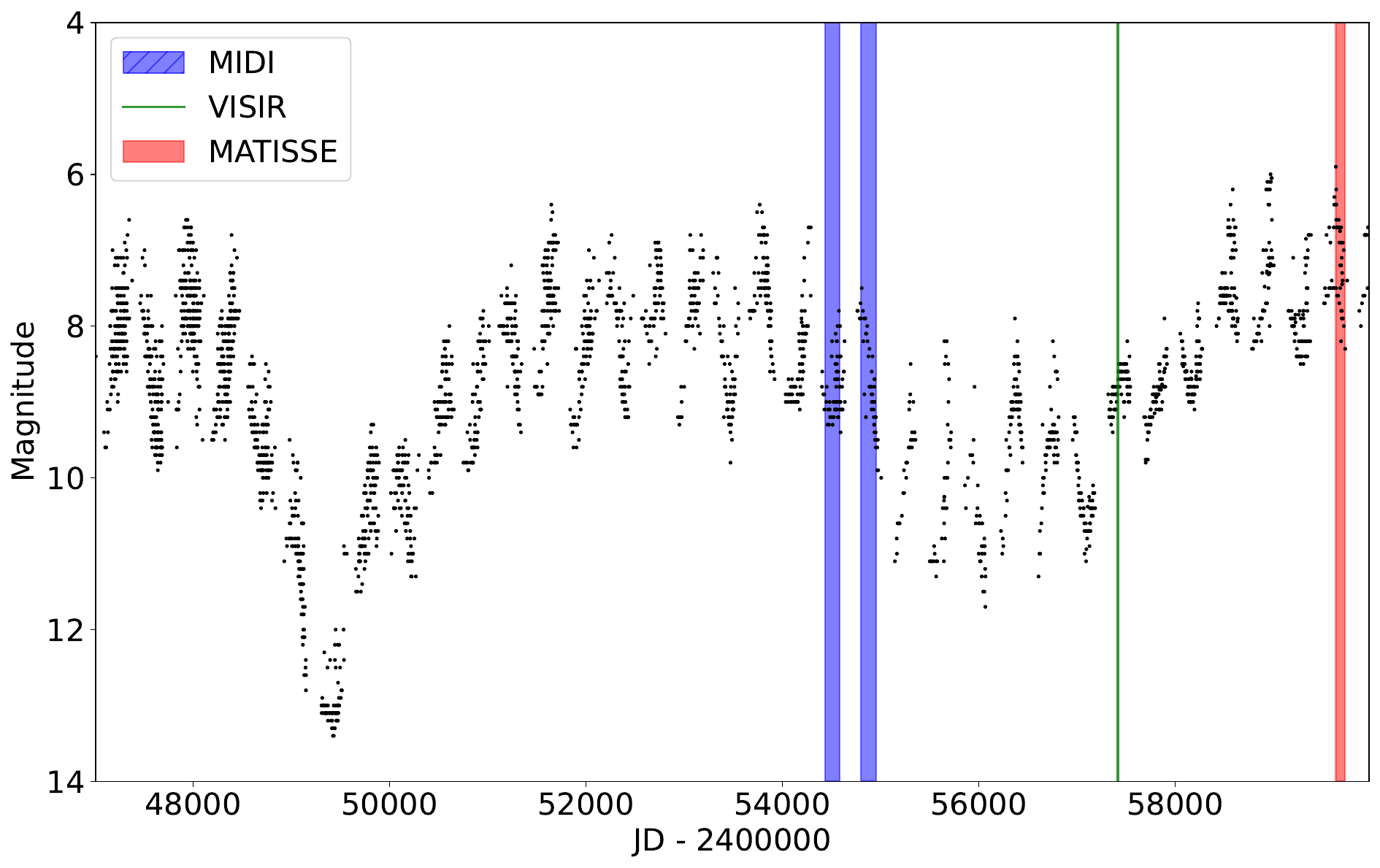}
     \caption{AAVSO light curve of V~Hya in the visual band (black points). The blue rectangle corresponds to the MIDI epoch of observation, the green line to the VISIR epoch, and the red filled rectangle to the MATISSE epoch.}
     \label{fig:light_curve}
 \end{figure}

V~Hya has been observed over the years by various instruments. \cite{Zhao_Geisler} presented VLTI/MIDI \citep{MIDI} data which are quite relevant for the current study. MIDI, the predecessor of MATISSE, was a 2-telescope beam combiner observing in the $N$-band with similar spectral setups as MATISSE.
The MIDI data from \cite{Zhao_Geisler} are in low spectral resolution (R~=~30), and they have been taken with various baselines and orientations 
as it can be seen in Fig. \ref{fig:UV}.
The observations were carried out between the end of 2007 and 2009, about 13 years before the MATISSE observations. In Fig. \ref{fig:light_curve}, the epochs of the observations are superimposed over V~Hya's visual light-curve from the American Association of Variable Star Observers (AAVSO, \citealt{AAVSO_ref}). Those data are used in the following sections to help the scientific interpretation of the MATISSE data.

As can be seen in Fig. \ref{fig:light_curve}, the observations taken by the three instruments sample different epochs. The respective 530~d pulsation and 17~yr obscuration phases ($\varphi_{\rm puls}$, $\varphi_{\rm obs}$) are (0.00$-$0.57, 0.32$-$0.34) for MIDI, (0.91, 0.73) for VISIR, and (0.10$-$0.27, 0.08$-$0.1) for MATISSE. The zero-phase of each cycle is taken as the closest light-maximum.

\section{Parametric modelling of the visibility}
\label{sect:parametricmodelling}
As a first step the interferometric data are interpreted using a parametric-model  approach. As the visibility curve is complex, a simple geometrical model (i.e. Gaussian, Uniform disc) cannot fit the data completely, however it provides a first characterisation of the general geometry of the object. GEM-FIND (a GEometrical Model Fitting for
INterferometric Data; \citealt{GEM-FIND}) fits the visibility data with composite geometrical models using a Levenberg-Marquardt minimization method (see  \citealt{Non_linear_least_square_2009ASPC..411..251M}). Five models with one or two components have been tested in this work: 1) an elliptical Gaussian, 2) an elliptical uniform disc, 3) a detached shell model composed of a uniform disc and a ring, 4) a binary model composed of two uniform discs, and 5) a combination of an elliptical Gaussian and a uniform disc of a given radius.  
GEM-FIND is designed for $N$-band analysis and was applied both to MIDI and MATISSE data. This tool was preferred over other available geometric-model tools because it was specifically designed to read text files such as the old MIDI pipeline output\footnote{The MIDI pipeline delivered also a first version of OIFITS, but they are not available for the V~Hya data.}.
The results, summarized in Table \ref{tab:GEMFIND-RHAPSODY}, are discussed in Sect. \ref{sect:visibility_fitting_results}.

The second tool used for parametric interpretation is RHAPSODY \citep{RHAPSODY}. RHAPSODY (Reconstructing Hankel rAdial Profiles in centro-Symmetric Objects with Discrete rings for astrophYsics) reconstructs the visibility with a set of concentric discrete uniform rings using a reduced chi-squared and Bayesian minimization method. The relative weight of the constraints from the measurements and the ones set by the prior is controlled by the so-called hyperparameter.Additionally, the method uses two angles of rotation (inclination and position angle) to add an angular dependency to the visibility profile, transforming the circular rings into elliptical ones.

The MATISSE closure phase is non-zero and non-$\pi$, implying strong asymmetries in the brightness distribution that cannot be modelled by fitting only the visibility curve. The non-centrosymmetric signature can be retrieved with image reconstruction methods that are able to simultaneously fit the visibility and the closure phase signal. 

\section{Image reconstruction procedure}
\label{sect:image_reconstruction}
The ($u,v$)-plane coverage obtained from our observations (see Fig. \ref{fig:UV}) is dense and spatially well distributed, making the MATISSE interferometric observables suitable for the use of image reconstruction techniques. The images were reconstructed using SQUEEZE \citep{SQUEEZE}, based on Simulated Annealing as minimization engine.
The images have been obtained by the following procedure: fifty chains are initialized on a different random image and run for 6000 iterations, using the transpectral regularizer. The final image is the mean image over the fifty chains. Images at one standard deviation from the mean display the same overall features (see Fig. \ref{fig:SQUEEZE_STD}).  
The image size is set to 128x128 pixels and the resolution is set to 1~mas for the $L$-band and 2~mas for the $N$-band. Four wavelength intervals, probing different molecular- and/or dust-forming regions have been selected to produce images (see Fig. \ref{figSED_DUSTY}): 3.15$-$3.20~$\mu$m (in $\rm C_2H_2$+ HCN absorption band), 3.60$-$3.65~$\mu$m (in the $L-$band pseudo-continuum), 10.50$-$10.55~$\mu$m (in the $N-$band pseudo-continuum), and 11.45$-$11.50~$\mu$m (near the SiC emission peak).
Images reconstructed on a narrow wavelength range ($\Delta \lambda$~=~0.05~$\mu$m) are needed to detect wavelength-dependent structural change in the source morphology. On the other hand, images reconstructed on a larger spectral band (here $\Delta \lambda$~=~1~$\mu$m) allow us to constrain with an increased precision geometrical features as more interferometric observables are fitted simultaneously. 
For the $N$-band, we also reconstruct images for the interval 10.50$-$11.50~$\mu$m, fitting simultaneously 2,016 square-visibility points and 1,343 closure phase points, for a final $\chi_r^2$ of 1.86.

To assess the reliability of the images obtained with SQUEEZE, two additional reconstruction packages have been used: MIRA by \cite{MiRA} and IRBis by \cite{IRBis}. The detailed procedure is described in Appendix \ref{appendix:image_reconstruction}. The final images obtained with the three methods are displayed in Fig. \ref{fig:compar_plot}, where they are convolved with a theoretical point spread function using a Gaussian with a full width at half-maximum (FWHM) equal to the interferometer resolution $\lambda/2B_{max}$. The images are found to be similar, revealing the convergence of the reconstruction process. 
As a last validation step, we performed a reconstruction with SQUEEZE on a simulated dataset to quantify the impact of the limited ($u,v$)-coverage on the image quality. The simulation step allows us to obtain an estimation of the feature arising from reconstruction artefacts. The results of the simulations are detailed in Appendix \ref{appendix:image_simulation}.

\section{Modelling the spectral energy distribution}
\label{sect:SED}
In this section we perform a qualitative study of the spectral energy distribution (SED) of V~Hya using a composite model, following the approach of \cite{Sacuto_2011A&A...525A..42S}.

Such analysis is performed in a first place to later interpret the MATISSE photometric and interferometric data and compare the model prediction with the infrared images.  
As the star undergoes a photometric variability (see Sect. \ref{sect:intro}), deriving precise stellar parameters from the fitting process of the optical region is out of scope with the available photometry. We therefore first select a carbon star hydrostatic model compatible with the properties of V~Hya from literature. Such synthetic spectrum is used as input radiation in the 1D-radiative-transfer DUSTY code to model the effect of a dusty environment on the spectral energy distribution \citep{DUSTY}.  

\subsection{Photometric data}
The calibrated MATISSE spectra are shown in the spectral energy distribution (Fig. \ref{figSED_DUSTY}) 
together with the HERSCHEL-PACS\footnote{\url{http://archives.esac.esa.int/hsa/whsa/}} spectrum and photometric data (listed in Table \ref{tab:Photometric_data})
, offering a spectral coverage from optical to far infrared.
To correct the photometric data from the interstellar reddening, the visual extinction equation is estimated using the tri-dimensional maps of the interstellar matter \citep{Lallement_2019A&A...625A.135L}
and the extinction curve of \cite{MW_exctinction_2009ApJ...705.1320G} were adopted. The extinction value in the visual is 0.09, and the impact of the interstellar extinction is negligible in the infrared. 

\subsection{Hydrostatic model}

To reproduce the stellar contribution to the total flux, we use the synthetic COMARCS\footnote{\url {http://stev.oapd.inaf.it/synphot/Cstars/}} spectra \citep{COMARCS}. The COMARCS spectra are computed from 1D hydrostatic models and take into account the absorption by atomic and molecular species from COMA (Copenhagen Opacities for Model Atmospheres) opacity tables. Further information about the COMA code and COMARCS spectra can be found in \cite{COMARCS} and \cite{COMA_2009A&A...494..403L}.
From the available COMARCS spectra, we select an input spectrum  compatible with the V~Hya inferred properties: temperature of 2700~K \citep{Knapp_1997A&A...326..318K}, a C/O ratio of 1.05 \citep{1986ApJS...62..373L} and a mass of 1~$M_{\odot}$ \citep{Kahane_1996A&A...314..871K}.
The other fixed parameters of the COMARCS model are a surface gravity $\log g$ of -0.4 and a solar metallicity. The final C/O ratio is 1.4 (see Sect. \ref{sect:the_dust_contribution}), well above the value 1.05 obtained by \cite{1986ApJS...62..373L}.

\subsection{The dust contribution}
\label{sect:the_dust_contribution}

To account for the dust contribution of the circumstellar environment, the 1D-radiative transfer code DUSTY\footnote{\url{http://faculty.washington.edu/ivezic/dusty_web/}} \citep{DUSTY} was used. Such 1D-approach constitutes a simplistic approximation but is justified by the fact that the MATISSE and VISIR images presented (see Fig. \ref{fig:visir_matisse} in Sect.~\ref{subsect:multiscale}) do not show strong asymmetric signatures which are instead observed at different wavelengths (see Sect.~\ref{sect:intro}). The main purpose of the DUSTY modeling is to provide a first approximation of the dust parameters and composition and to calibrate the flux of the images. A multi-dimensional modeling of the system is beyond the scope of this paper.

DUSTY solves the radiative transfer problem in a circumstellar environment under the  spherical-symmetry assumption.
For AGB stars, the density distribution is assumed to be driven by the pressure on dust grains which produces winds, known as radiatively-driven winds \citep{Hofner_Olofsson_2018A&ARv..26....1H}. The dust grain-size distribution is assumed to follow the standard MRN 
power law \citep{MNR_1977ApJ...217..425M} for grain radius, $a$, ranging from 0.005~$\mu$m to 0.25~$\mu$m. The dust shell density is assumed to vary as $\rho(r) \propto r^{-2}$.
For carbon stars, the grain mixture is expected to be mainly composed of amorphous carbon, AmC \citep{AmC_1991ApJ...377..526R}, with a small contribution of silicon carbide, SiC \citep{SiC_1988A&A...194..335P}.

To derive the uncertainty of the input parameters AmC/SiC ratio, inner radius, and shell thickness, we assessed their impact on the SED. 
The adopted AmC/SiC ratio is 90/10, and a variation by $\pm5$\% of the SiC fraction does not allow us to reproduce the shape of the 11.3~$\mu$m SiC feature. 
The inner radius of the circumstellar envelope (CSE) is upper bounded by the location of the dust grain temperature of $T=1300$~K, as 
found for IRC +10216 \citep{1994AJ....107.1469D}. 
To correctly fit simultaneously the $L$- and $N$-band fluxes, an inner boundary located at 20~au is required, corresponding to an inner boundary temperature of 1050~K. Temperatures above 1200~K would lead to a lack of flux in the MATISSE bands while temperatures below 950~K imply a flux excess in the N$-$band compared to the L$-$band. The value adopted for the temperature at the inner CSE boundary (thus fixing correspondingly the value of the inner radius) was taken in between the two threshold values mentioned above. 

The geometrical thickness of the CSE was arbitrarily set to 10,000 dust-shell inner radii. Values ranging from 100 to $10^5$ have no significant impact on the SED for wavelengths below 100 $\mu$m \citep{Sargent_2010ApJ...716..878S}. 
Finally, the C/O ratio of the COMARCS model was fine-tuned to reproduce the absorption feature around 3.15 $\mu$m.

\begin{table}[t]
    \centering
        \caption{Dusty circumstellar envelope parameters for V~Hya obtained with DUSTY. }
    \label{tab:DUSTY_table}
    \begin{tabular}{lr}
    \hline
    \hline
    Parameter & Value \\
    \hline
      Chemical composition & 90 $\pm$ 5\% AmC  \\
      & 10 $\pm$ 5\% SiC\\
      Grain distribution & $n(a)\propto a^{-3.5}$\\
      Shell relative thickness, $Y$ & 10000\\
      Inner radius, $R_{\rm in}$~[au]  & 20 $\pm$ 2\\
      Inner boundary temperature, $T_{\rm in}$~[K]   & 1050 $\pm$ 90\\
        \hline
    \end{tabular}
\end{table}

\begin{figure}[t]
    \centering
    \includegraphics[width = 0.5\textwidth]{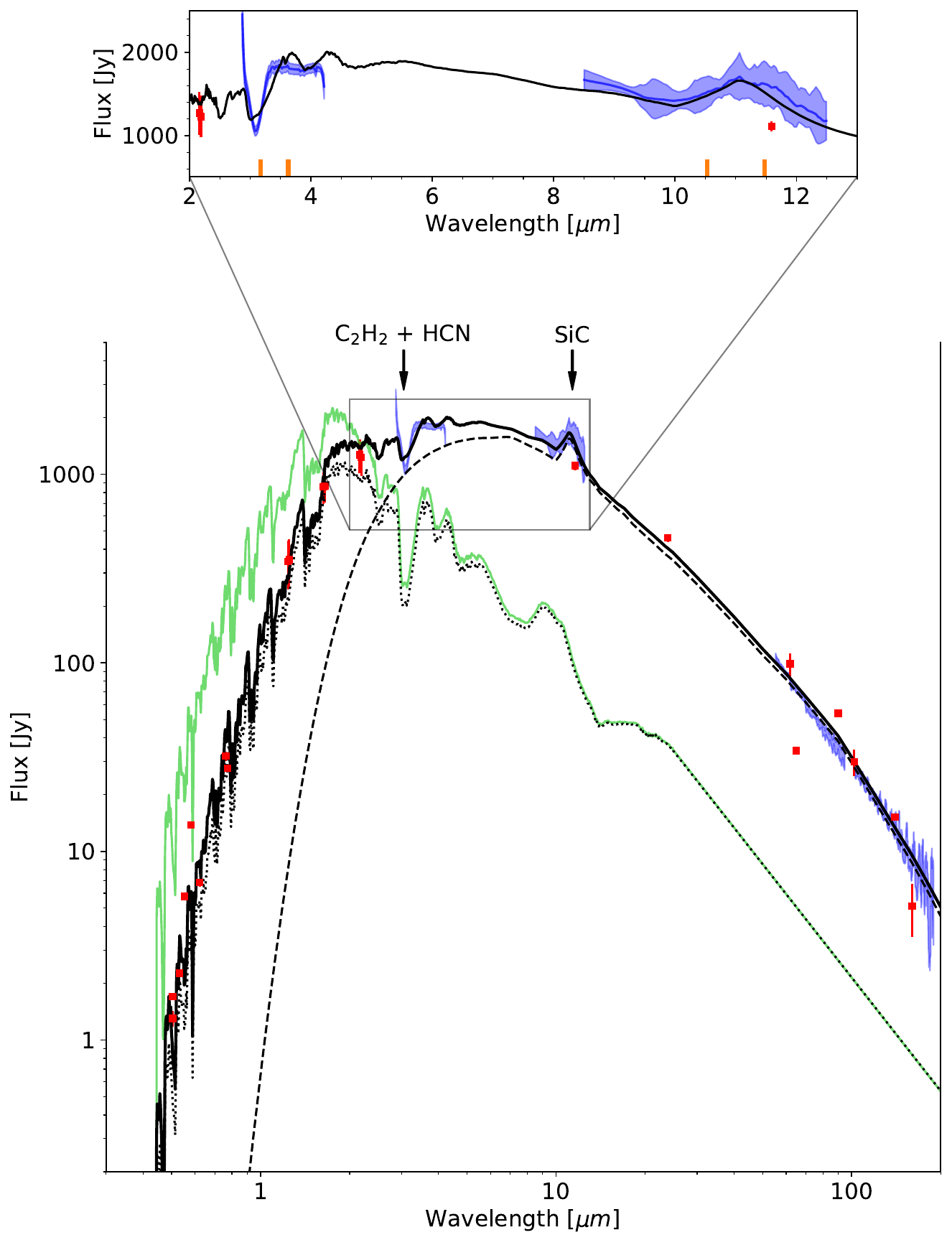}
    \caption{SED of V~Hya. The red dots represent the photometric data from Table \ref{tab:Photometric_data}. The blue shades are, from left to right, the MATISSE ($L$- and $N$-bands) and the HERSCHEL-PACS spectrum. 
    The green line is the (unreddened) COMARCS stellar spectrum. The dotted line represents the reddened spectrum. The dashed line is the contribution of the dust emission and the thick black line represents the resulting DUSTY model. 
    The inset zooms onto the MATISSE spectra. The orange vertical bars mark the wavelength region selected for image reconstruction.}
    \label{figSED_DUSTY}
\end{figure}

The composite COMARCS+DUSTY model is displayed in Fig.~\ref{figSED_DUSTY}, together with the photometric data. 
The bolometric flux obtained is 3$\times10^{-9}$~W/m$^2$, corresponding to a stellar luminosity of $L$ = 18\;000~$\rm L_{\odot}$ for the adopted distance of 434~pc.


Longward of 1.7 $\mu$m, the dust emission becomes the main contributor to the overall flux and dominates in the thermal infrared range ($L$- and $N$-bands).


\section{Mid-infrared results}
In this section we first present the results of the parametric modelling of the visibility curves for the MATISSE and the MIDI data. Then we present the reconstructed images obtained for the MATISSE data. The latter are then used to asses the temporal variability in the visibility data between the two epochs. Finally, the MATISSE $N$-band image is flux-calibrated and compared with the VISIR coronagraphic image to get large-scale information of the CSE.

\subsection{Visibility fitting}
\label{sect:visibility_fitting_results}
The results of the parametric modelling for MATISSE and MIDI are summarized in Table \ref{tab:GEMFIND-RHAPSODY}. 

\begin{table}[t]
    \begin{center}
    \caption{Results of the visibility fitting for the best GEM-FIND model and RHAPSODY reconstructed intensity profile.}
    \label{tab:GEMFIND-RHAPSODY}
    \begin{tabular}{lrr}
    \hline
    \hline
       \textbf{GEM-FIND}& MIDI & MATISSE \\
              \hline
       $\lambda$~[$\mu$m]& 11.35 & 11.35\\

        PA [$^\circ$]   & 160 $\pm$ 10 & 71 $\pm$ 20\\
        Axis ratio & 0.7 $\pm$ 0.1 & 0.8$\pm$ 0.2\\
        Brightness ratio & 4.9 $\pm$ 0.1 &  5.5 $\pm$ 0.3     \\
        FWHM [mas] & 21.3$\pm$ 0.2 & 16.4 $\pm$  0.5 \\
        $\chi^2$ & 29 & 3.68\\
            \hline
    \hline
        \textbf{RHAPSODY} &  MATISSE  & MATISSE \\
        \hline
         $\lambda~[\mu$m]& 3.66 & 11.35\\

        PA [$^\circ$] &150 &65\\
        Axis ratio & 0.5&0.7\\
        Hyperparameter& $10^4$& $5\times10^4$\\
        $\chi^2$ &2.18 & 2.74\\
     \hline
    \end{tabular}    
    \tablefoot{
        PA is the position angle, counted from North to East. The brightness ratio is between the uniform disc and the Gaussian.
     }
     \end{center}
\end{table}
The best GEM-FIND model for both MATISSE and MIDI is composed of an over-resolved uniform disc (with a diameter fixed to 500~mas to mimic a background component) and an elliptical Gaussian. The extended uniform disc contributes to 80\% of the total flux. 
Those values are consistent with the low visibility observed at small baselines (Fig. \ref{fig:vis2t3_all}), revealing the presence of an extended object fully resolved by the instrument.
Such ratio is similar to that previously found by \cite{Zhao_Geisler} using circular models (a uniform disc and a Gaussian) to fit the MIDI data: a central source of radius 19.3~$\pm$3.5~mas contributing to 23\% of the total flux. 
The FWHM of the central elliptical Gaussian, its brightness ratio and its axis ratio, listed in Table~\ref{tab:GEMFIND-RHAPSODY}, reveal similar values for both MIDI and MATISSE data. The elongation is well pronounced, resulting from the position-angle dependence of the visibility. Remarkably, the position angle (PA, counted from North to East) of the major axis obtained is shifted by about 90$^\circ$ between the two datasets. This change in the ellipse orientation between the two epochs is discussed in Sect.~\ref{sect:discussion}. 
The reconstructed intensity profile obtained with RHAPSODY at 11.35~$\mu$m and the model obtained with GEM-FIND for the same wavelength exhibit the same geometry: an elongated structure in the North-East direction with an axis ratio of 0.7 (Table~\ref{tab:GEMFIND-RHAPSODY}). 
In the $L$-band at 3.66~$\mu$m, the elongation is perpendicular to the one obtained in the $N$-band. 

In summary, the geometry of the central source presents a position-angle dependence but the two infrared bands do not share the same PA for their elongation axis. For the $N$-band, the PA is not stationary on a 13-yr timescale. The temporal evolution of the $N$-band observables is further discussed in Sects.~\ref{subsect:multiepoch} and \ref{subsect:impactofcompanion}.

\subsection{MATISSE images}
\begin{figure*}[h]
    \includegraphics[width = \textwidth]{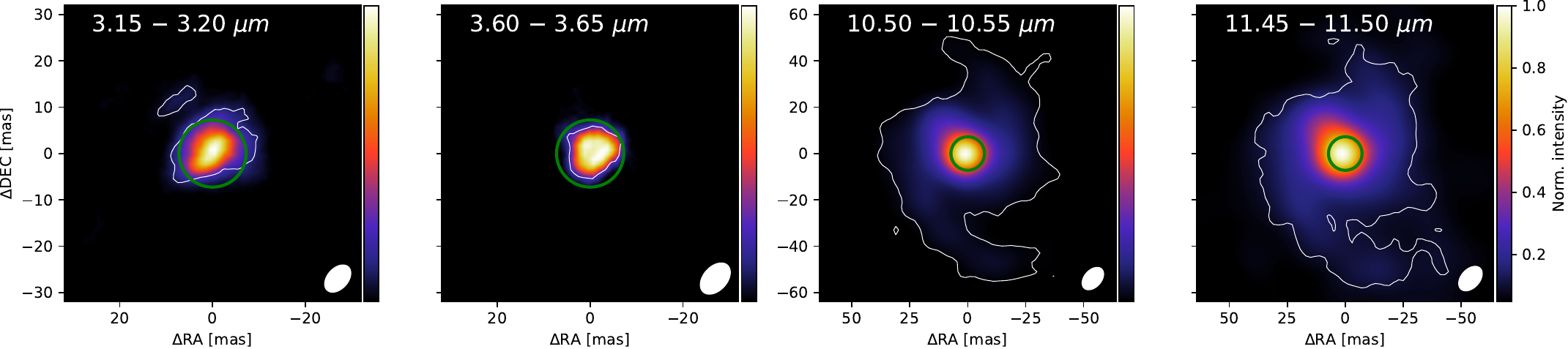}
    \caption{MATISSE images of V~Hya reconstructed with SQUEEZE at four different spectral ranges across the $L$- and $N$-bands. The green circle represents the expected star diameter of 14.5~mas as measured in the $K$-band \citep{IOTA_interf}. The white contour is drawn at 3$\sigma$ (as described in Appendix \ref{appendix:image_reconstruction}). 
     The white ellipse in the bottom-right corner represents the mean synthesized primary beam.}
    \label{fig:LMN}
\end{figure*}

The images obtained with SQUEEZE at four selected wavelength ranges are shown in Fig.~\ref{fig:LMN}. Their corresponding visibility and closure phase fits are displayed in Fig.~\ref{fig:vis2t3_all}. 

\subsubsection{$L$-band}
\label{subsection:Lband_results}
The image at 3.15$-$3.20$~\mu$m (leftmost panel of Fig.~\ref{fig:LMN}) probes the $\rm{C_2H_2}$ and HCN absorption features. At those wavelengths the image obtained shows an elongated central source with an extension visible in the north-east quadrant. 
At 3.60$-$3.65$~\mu$m (second leftmost panel of Fig.~\ref{fig:LMN}), falling within the pseudo-continuum (see the inset in Fig.~\ref{figSED_DUSTY}), the central source 
does not exhibit a circular shape either: an hook-like extension is directed toward the north-east quadrant and the central source diameter is smaller than the estimation from the K-band \cite{IOTA_interf}. 
The interpretation of these features is presented in Sect.~\ref{sect:discussion}.

\subsubsection{$N$-band}
\label{subsection:Nband_results}
The $N$-band images at 10.50$-$10.55$~\mu$m and 11.45$-$11.50$~\mu$m  (Fig.~\ref{fig:LMN}, right panels) are displayed with a field-of-view twice larger to fully characterize the circumstellar environment.
Both images exhibit a similar geometry: a central source of diameter about 14~mas (enclosed in the green circle, and corresponding to a fractional intensity of 0.7) and an extended emission in the North-East quadrant, hereafter referred as the "20~mas extension", whose PA is consistent with the elongation axis found from our preliminary model fitting (see Table \ref{tab:GEMFIND-RHAPSODY}). 
Comparing the angular extension of the circumstellar dust in the two rightmost panels of Fig.~\ref{fig:LMN} reveals that the source appears more compact at 10.50~$\mu$m than at 11.45$~\mu$m.

\begin{figure}[t]
    \centering
    \includegraphics[width = 0.5\textwidth]{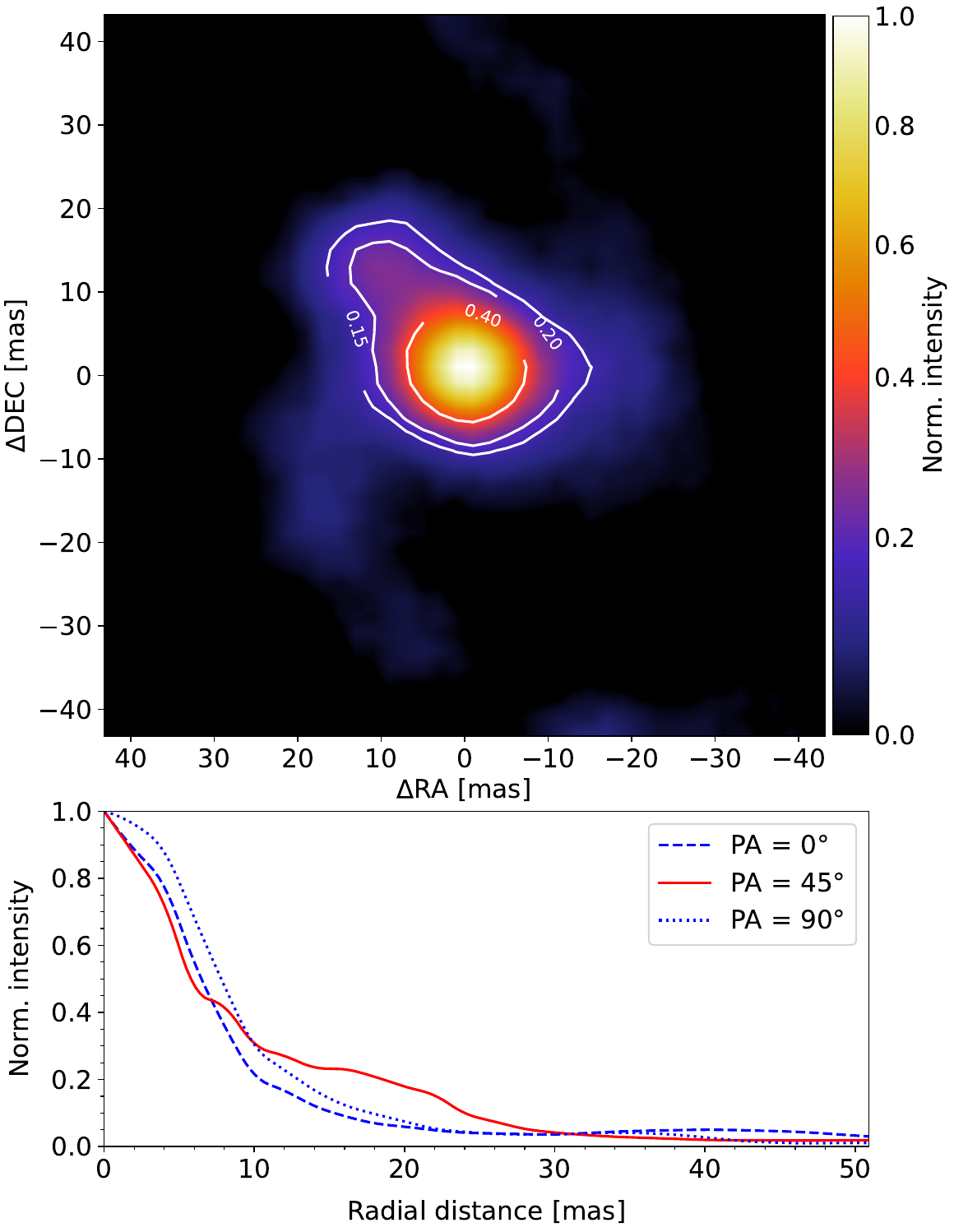}
    \caption{Top: MATISSE image of V~Hya reconstructed with SQUEEZE for the spectral interval 10.50-11.50~$\mu$m. The white contour levels, labelled with the fraction of the peak intensity, highlight the elongation towards the upper left corner. In the text,
    this feature is referred to as the "20 mas extension". Bottom: Radial cuts (starting from the central brightest pixel) in the image for  different PAs, as indicated.}
    \label{fig:dusty_blob_radial}
\end{figure}

The image in the 10.50$-$11.50$~\mu$m band displayed in Fig.~\ref{fig:dusty_blob_radial}, together with the contour levels and the radial cuts at PAs of 0$^\circ$, 45$^\circ$, 90$^\circ$, illustrate the strong angular asymmetries
(Fig.~\ref{fig:RGB_image}, discussed in Sect.~\ref{subsect:impactofcompanion} below, presents the same data, but in the form of a $L$- and $N$-band colour-composite image).

Given the simulations described in Appendix~\ref{appendix:image_simulation}, we may state that the spiral-like structure appearing in the $N$-band (best visible in the rightmost panel of Fig.~\ref{fig:LMN} and to a lesser extent in Fig.~\ref{fig:dusty_blob_radial}) is likely an artefact of the image reconstruction process related to the specific ($u,v$)-plane coverage.

\subsection{Multi-epoch analysis of the visibility curve}
\label{subsect:multiepoch}
\begin{figure}[t]
    \includegraphics[width = 0.49\textwidth]{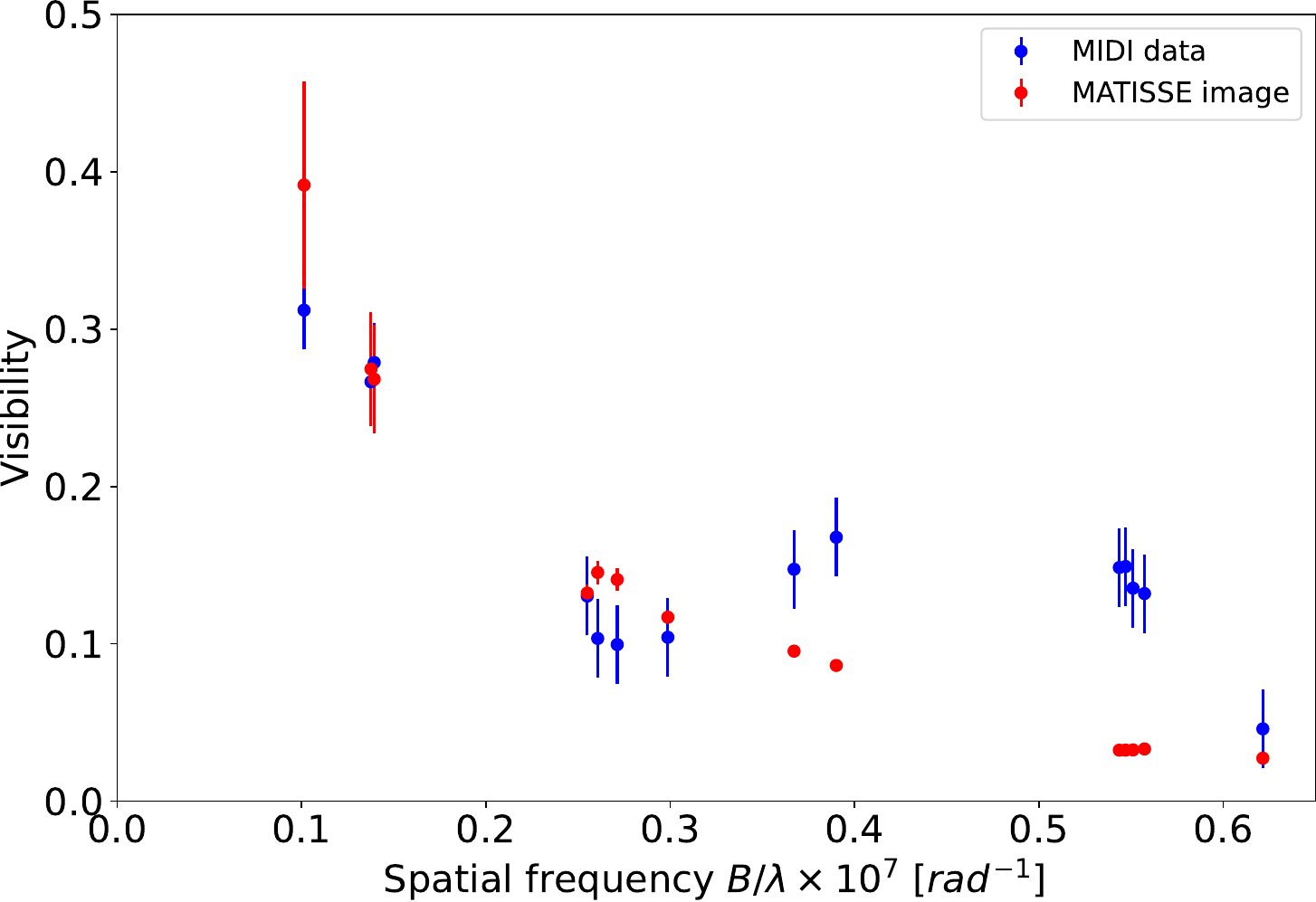}
    \caption{Visibility measurements at 11.45~$\mu$m at phase $\phi_{\rm puls}$ = 0.18$\pm$0.08. The blue points are the MIDI measurements from \cite{Zhao_Geisler}. The red points are extracted from the MATISSE reconstructed image, at the ($u,v$)-coordinates of MIDI (see Fig. \ref{fig:UV}). }
    \label{fig:MIDI-MATISSE}
\end{figure}

The MIDI and MATISSE observations probe the same wavelength band but are separated in time by about a decade (see Fig.~\ref{fig:light_curve}). Their ($u,v$)-plane coverage being slightly different in terms of PA or baseline lengths (see Fig.~\ref{fig:UV}), a direct comparison of the visibility measurements is not straightforward. In Sect.~\ref{sect:visibility_fitting_results}, we fitted the visibilities using simple geometrical models and showed that the best-fitting models are not oriented along the same direction at the two epochs. 
To ensure that this difference is not an artefact of the different $(u,v)$-coverages, we extract from the MATISSE reconstructed image the visibilities at the exact ($u,v$)-coordinates of the MIDI observations.
To eliminate the effect of pulsation, only the MIDI observations matching the pulsation phase of MATISSE ($\phi_{\rm puls}$ = 0.10$-$0.27) are selected.
The extrapolated visibility curve obtained from the MATISSE $N$-band image and the MIDI data are displayed in Fig.~\ref{fig:MIDI-MATISSE}. The uncertainties on the synthetic MATISSE visibilities are extracted from the MEAN$\pm$STD image displayed in Fig.~\ref{fig:SQUEEZE_STD}. 
The general shape of the visibility curve strongly differs between the two epochs and is unlikely to originate from the flux variation induced by the pulsation.

\subsection{Multi-scale analysis of the CSE}
\label{subsect:multiscale}

\begin{figure*}[t]
    \centering
   \caption{VISIR coronagraphic images (left) and the corresponding MATISSE reconstructed image (right). The scale of the MATISSE field of view is represented by a white rectangle at the center of the VISIR image. The green continuous circle represents the size of the stellar disc as measured in the $K$-band while the green dashed circle is its interpolated value in the $N$-band (see text). North is up, East is left.}
    \label{fig:visir_matisse}
\end{figure*}
VISIR and MATISSE probe similar wavelength ranges but at different spatial scales. 
Figure~\ref{fig:visir_matisse} displays the coronagraphic image obtained with VISIR, together with the MATISSE $N$-band image at 11.45$-$11.50~$\mu$m. The field of view of VISIR is about 10 times larger than the MATISSE one. The VISIR image displays a circular emission pattern. 
The overall integrated flux (without coronagraph) is equal to 1756 Jy, consistent with the MATISSE photometry (see Fig.~\ref{figSED_DUSTY}).
In the following we describe the procedure applied to calibrate MATISSE images in flux, as well as the comparison of the radial profiles of the images obtained from the two instruments. 

As the MATISSE image is reconstructed from visibility and closure-phase measurements, each pixel value corresponds to its relative weight over the image and does not bear information about the absolute intensity of the source. 
To flux calibrate the image, we use as conversion factor the stellar flux from the SED computation  (see Fig.~\ref{figSED_DUSTY} where the attenuated stellar radiation at 11.45~$\mu$m is 102~Jy). 
In the MATISSE image, only the pixels inside the photospheric radius contribute to the flux of the central component.
The $N$-band uniform disc radius is derived by means of dynamic model predictions from the measured $K$-band value (7.25~mas from \citealt{IOTA_interf}).
A ratio of about 1.6 between $K$- and $N$-band radii is expected for a carbon star with a stellar wind \citep{Paladini_model_radius_2009A&A...501.1073P}, leading to an $N$-band radius of 12.3 mas.
The $K$- and $N$-band radii are represented by the continuous and dashed green circles in Fig.~\ref{fig:visir_matisse}. 
The relative flux contribution of a central disc of 12.3~mas radius is 11\% and corresponds to the stellar contribution plus an additional contribution from the dust emission located in the line-of-sight. The dust contribution is corrected by removing the contribution of an annulus of half\footnote{This factor 1/2 comes from the fact that we need to subtract the contribution of the dust located in front of the stellar disc using an annulus around the stellar disc which includes dust located both 'in front' and 'behind' the star along these grazing lines of sight. Assuming that the dust distribution is symmetric 'in front' and 'behind' the star justifies the factor 1/2, along with the fact that the dust is optically thin at the considered wavelengths (see Sect. \ref{sect:the_dust_contribution}).} the stellar area located just around the star radius.
After implementing this correction, the stellar contribution drops down to about 8\% of the image flux density.

\begin{figure}[t]
    \centering
   \includegraphics[width = 0.5\textwidth]{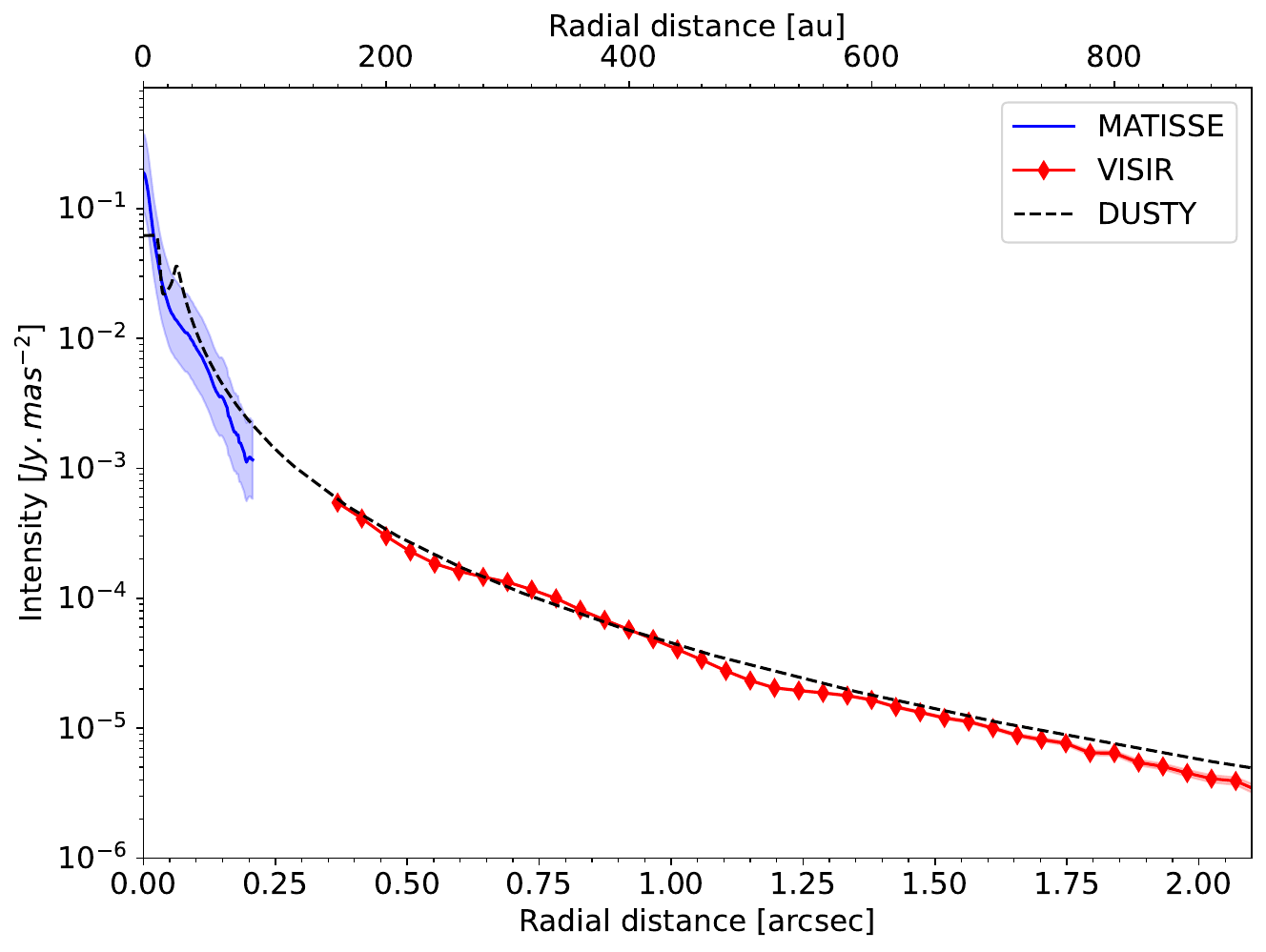}
    \caption{Radial intensity cut of the 11.4 µm emission around V~Hya, averaged over azimuthal angle. The MATISSE radial profiles (blue line, and its relative error represented by the shaded light blue area) are calibrated taking a photospheric radius of 12.3$\pm$5.0 mas. The VISIR radial profiles are displayed with red square symbols. The DUSTY model profile, downgraded to the MATISSE resolution, is displayed by the dashed black line.}
    \label{fig:radial_cut}
\end{figure}

The radial intensity profile (averaged over the azimuthal angle) for the MATISSE $N$-band and VISIR images is displayed in Fig. \ref{fig:radial_cut}. 
The two profiles are shown together with the radial profile from the DUSTY modelling (see description in Sect. \ref{sect:the_dust_contribution}) for the same wavelength range. The good agreement between the model and the composite VISIR-MATISSE radial profile implies that the dust density profile roughly follows the $r^{-2}$ trend adopted in the DUSTY model computation.

\section{Discussion}
\label{sect:discussion}
In this section, we interpret the features found in the MATISSE images, and discuss their properties and possible origin. We confront our interpretation with the signature predicted for a close-by companion; finally we discuss the overall shape of the dusty environment including the larger-scale image obtained from VISIR (see Fig. \ref{fig:visir_matisse}) and observations previously reported in the literature. 
\subsection{The morphology}
\label{subsect:Morphology}
\subsubsection{Asymmetric stellar photosphere}
As mentioned in Sect. \ref{subsection:Lband_results}, the images in the $L$-band (Fig.~\ref{fig:LMN}) display an elongated structure (with a major axis oriented at about $150^\circ$, counted from North to East). 
Given the spatial scales of the asymmetries, they are likely to originate from the processes in the photosphere: the asymmetries are found within the expected angular radius of 7.25~mas (cf. Fig. \ref{fig:LMN}). 
It is known from modelling that the dynamic process of pulsation and convection taking place in AGB stars can lead to surface-brightness asymmetries due to the presence of hot and cool convective cells on the photosphere (\citealt{2011A&A...528A.120C}, \citealt{2018Natur.553..310P}). Previous near-IR interferometric observations of the carbon-rich star \object{R Scl} show that the dimension of those cells can be up to 1/3 the scale of the stellar photosphere with a brightness ratio of 1:2.5 \citep{RScl_Wittkowski2017A&A...601A...3W}.

The shape of the stellar image in the $L$-band also 
displays an extension in the North-East direction (see Fig.~\ref{fig:LMN}). Such structure could as well be related to the underlying convective atmosphere, like the one observed in \object{R Scl} \citep{2022A&A...665A..32D}, however in Sect.~\ref{subsect:impactofcompanion} we speculate that it could be linked to the binary companion.

\subsubsection{North-East thermal emission}
\label{subsection_discussion_SW_emission}
The images in the $N$-band also display an elongation. However, contrary to the $L$-band, the major axis is oriented along the direction corresponding to the position angle of 45$^\circ$, and is therefore perpendicular to the $L$-band major axis. As described in Sect.~\ref{subsect:multiscale}, the central emission is composed of a circular disc with a radius of $\sim$12~mas and a 20~mas extension in the North-East direction. The extension is more prominent at longer wavelengths. The SED modelling (see Sect.~\ref{sect:SED}) shows that the $N$-band range is dominated by the dust emission. The dusty envelope is mainly composed of amorphous carbon, and silicon carbide is responsible of the emission feature seen around 11.4~$\mu$m. As a consequence, the extension seen is likely to be related to the dust-forming region.

Two scenarios can be proposed to explain the origin of  a dust clump located around the AGB star. Either an asymmetric ejection of material during the pulsation that cools down when pushed away from the AGB star, creating dust emission away from the central body in a random direction, as modelled by \citet{Suzanne_ref_2019A&A...623A.158H} and \citet{Freytag2023}.
Or, the other explanation that we tend to favour given the binary nature of the star (see Sect.~\ref{sect:intro}) would be the presence of a dust clump induced by the presence of the close companion. \citet{Mohamed2012BaltA..21...88M} and \citet{Chen2020} (to cite just a few) modelled the outflows from detached Mira-type binaries as "wind Roche-lobe overflow" (WRLOF). The simulations lead to the creation of complex and asymmetric structures shaped by the binary motion, with particle enhancement in the vicinity of the companion location. 
In the next section, we argue why the binary scenario is considered as more likely and confront the interferometric observables with the orbital model.

\begin{figure}[t]
    \includegraphics[width = 0.5\textwidth]{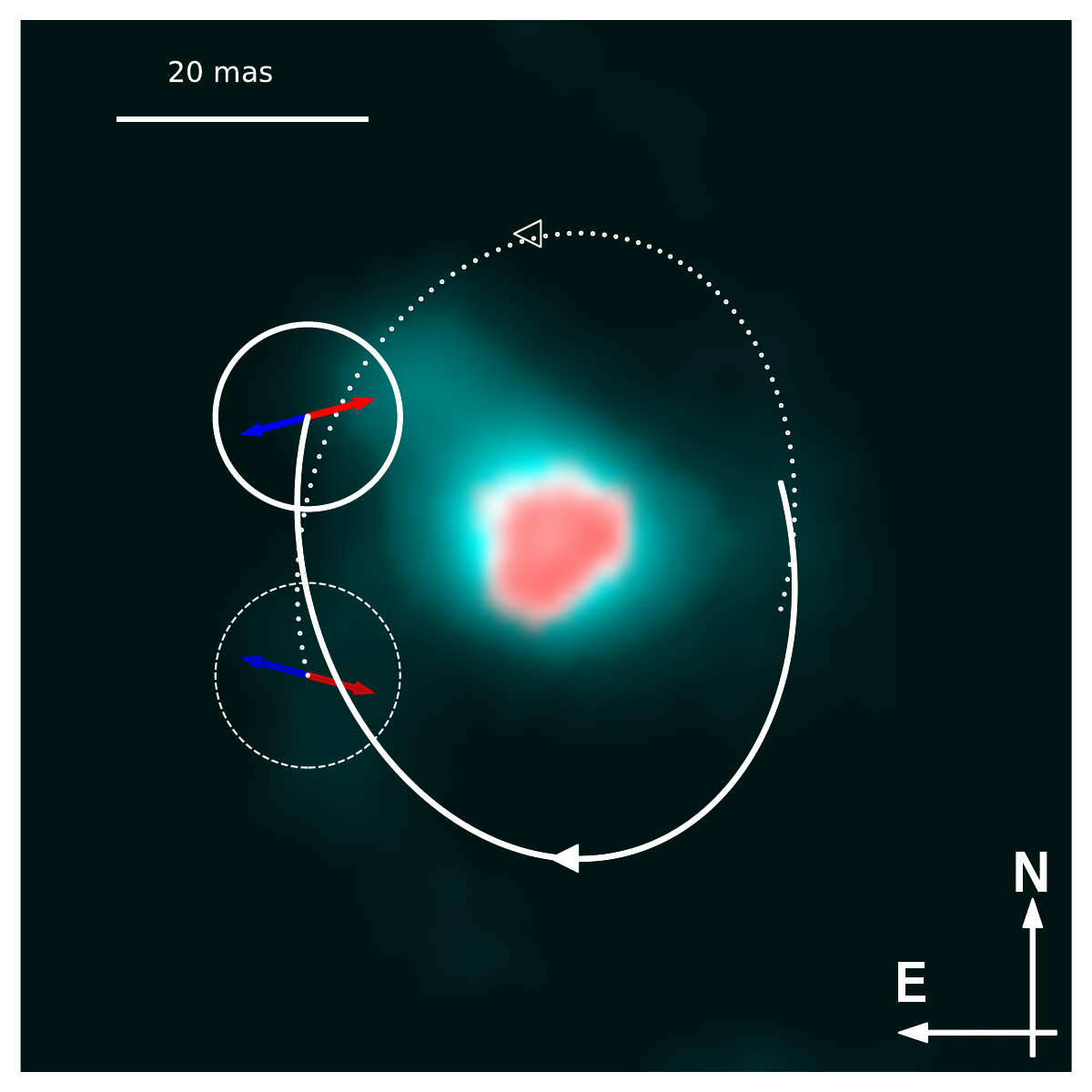}
    \caption{Composite colour image with flux in $L$-band (3.60~$\mu$m) displayed in red, and flux in $N$-band (10.50-11.50~$\mu$m) in blue. The white solid (respectively dashed) circle represents the 3$\sigma$-position prediction at MATISSE epoch for the clockwise (resp. counter-clockwise) orbital motion. The white arcs show the path of the companion around the primary since the last superior conjunction while the blue (resp. red) arrow represents the sky-projected jet axis for the blue- (resp. red-)shifted direction.}
    \label{fig:RGB_image}
\end{figure}

\subsection{The impact of a companion}

\label{subsect:impactofcompanion}

\cite{self_citation} proposed for V~Hya an orbital solution with a period of 17.5 years (compatible with the light-curve modulation), a low eccentricity, an inclination of about 40$^\circ$, and a semi-major axis  of about 11.2~$\pm$~1.5~au (25.8~$\pm$~3.5~mas at 434~pc), 
hence the companion is expected to fall in the MATISSE field of view.
More precisely, knowing that the orbital phase difference between the system superior conjunction (corresponding to the last dimming event) and the MATISSE observation is $\Delta \phi=0.6$, it is possible to estimate the companion position in 2022. The companion (V~Hya B) is expected to be offset from the central source by about 20~mas with a position angle of $60^\circ\pm20^\circ$ or $120^\circ\pm20^\circ$, depending on whether a clockwise or counterclockwise motion around the AGB star is adopted \citep[see Sect. 3.2 in ][]{self_citation}. These two possible positions predicted for the companion at the epoch of the MATISSE observation, and the corresponding orbital motion since the last superior conjunction are displayed on the colour-composite MATISSE image (Fig.~\ref{fig:RGB_image}). The North-East 20~mas extension, agrees with the prediction of the companion position for a clockwise orbital motion.

The clockwise orbital motion is also slightly preferred by the difference in the proper motions of V~Hya between the latest two Gaia data releases (GDR2 and GDR3 at epochs 2015.5 and 2016.0 respectively). The proper motion curve in declination between the two epochs can be described by a linear trend with a negative slope (non-zero within a 75\% confidence interval). Such a proper motion trend around the GDR2 and GDR3 epochs in turn corresponds to a declination sky-motion of V~Hya following a downward-facing parabola, hence an opposite (upward-facing parabola) for the companion. This condition is met by the clockwise path in Fig. \ref{fig:RGB_image}.

Additionally, the relation between the 20~mas extension and the companion orbital motion is supported by the PA of the elongation 
found in the earlier MIDI data. The MATISSE and MIDI data sets are separated by about 13 years and, given the orbital period of 17.5 years, this corresponds to a phase difference of 0.75. Considering that a full orbital revolution would induce a complete revolution of the 20~mas extension, the PA offset between the two epochs separated by 13 years would be equal to 270$^\circ$, in accordance with the PA values found by the model fitting (see Table~\ref{tab:GEMFIND-RHAPSODY}).

The above arguments tend to favour the position prediction that matches the extension seen in the MATISSE $N$-band image. 
More precisely, since the companion is expected to be a main-sequence star, necessarily hotter than the AGB star \citep{Sahai_2008ApJ...689.1274S}, it is supposed to be unresolved and not directly detectable in the infrared band (about a thousand times fainter than the AGB star). The infrared emission observed close to the companion position would instead reveal the presence of particle (dust and molecules) enhancement.
\cite{Chen2020} predict particle enhancement in the vicinity of the companion location, up to a 3 orders of magnitude above the ambient density.

The hook observed in V~Hya $L$-band images (Fig.~\ref{fig:LMN}) is pointing toward the unseen companion located North-East from V~Hya (and 7~au further away from the observer than the primary), suggesting the presence of a possible accreting flow, as imaged for Mira in the UV \citep{Mira_AB_2004RMxAC..20...92K}.
The North-East emission prominent in the $L$-band image taken in the $\rm C_2H_2$ absorption feature (Fig.~\ref{fig:LMN}, first panel) but not in the $L$-band image taken in the pseudo-continuum (Fig.~\ref{fig:LMN}, second panel) could correspond to a $\rm C_2H_2$ enhancement close to the companion - $\rm C_2H_2$ being a building block in the formation of amorphous carbon \citep{refSiC_1988ioch.rept.....H} and SiC \citep{RefSiC_2008A&A...483..661P}. Or it could be an opacity effect that reduces the stellar contribution to the total flux in the molecular absorbing feature (see Fig.~\ref{sect:SED}) and therefore increase the relative intensity of the clump.

The remaining open question concerns the structure (geometry and composition) of the clump and its possible connection with the periodic obscuration. The MATISSE $N$-band image confirms the presence of such localised emission, but, as the image only gives us the sky-projected geometry of the emission region (seen as a 20~mas extension), no 3D-geometry can be retrieved.

\subsection{Link with the periodic obscuration}
 Constraints on the dust distribution in the system come from the light-curve shape and the system inclination: the dust clump should be opaque and extended along the jet axis to obscure the primary for about one third of the orbital cycle and to cause up to $\Delta m=3$~mag obscuration in the optical during the superior conjunction (see Fig.~\ref{fig:light_curve} and \citealt{self_citation}).
The location of the obscuring cloud, above the orbital plane, cannot be easily explained by WLROF-type mass transfer but seems to require more sophisticated models involving jets, of the kind encountered in pre-main sequence stars and post-AGB binaries (\citealt{Ferreira_2007LNP...723.....F}, \citealt{Jet_post_AGB_2024arXiv240116370V}). 

The mass grain column density $n$ required for the obscuration can be estimated through $\Delta m = 1.086 \; n \; \kappa_{\rm ext}$, where $\kappa_{\rm ext}$ is the
total extinction cross section per unit mass in the visible range. 
Assuming Rayleigh regime, $\kappa^\lambda_{\rm ext}$  is given by $Q/a \times (3/4\rho)$ = 3$\times \rm 10^{4}$~cm$^{2}$~g$^{-1}$, where $\rho$ is the adopted mean grain density of 1.85 g~cm$^{-3}$ \citep{AmC_1991ApJ...377..526R}, and $Q/a$ (the extinction efficiency over the grain radius) is about $10^5$~$\rm cm^{-1}$ at 0.55~$\mu$m \citep{ref_QA_1999A&A...349..243A}. Hence, a mass column density of the order $\rm 10^{-4}$ g~cm$^{-2}$ would be required.
This implies a total dust mass (contained in the absorbing cylinder located in the line of sight with an area subtended by 
V~Hya photosphere with radius of 7.25~mas) of the order of $10^{-10}$~M$_{\odot}$. 
During the dimming events, this amount of dust would alter the SED shape by inducing an additional absorbing source in the optical to near-infrared region.

The dust dynamics (dust cloud continuously formed or attached to the companion) cannot be retrieved with a single image of the dust-forming region. The only conclusion that can be drawn at this point is that a particle enhancement around the companion is present and that the obscuring dust clump should extend along the jet axis.
 Further observations in the infrared $N$-band taken at a different orbital phase (e.g., during a dimming event) would constrain the geometrical structure and extend of the "20-mas extension" by probing the component positions from a different vantage point. 

\subsection{Large-scale morphology}
\label{subsect:discussion_largescale}

The largest structure resolved by the MATISSE images is limited by the shortest baseline ($\sim$15 m) of the interferometric measurements. As a consequence, any structure more extended  than the largest recoverable scale ($\lambda/B_{\rm min} \sim 165$~mas at 12 $\mu$m) is not resolved. Both the low visibilities measured at the shortest baseline separation (see Appendix, Fig.~\ref{fig:vis2t3_all}) and the SED (Fig~\ref{figSED_DUSTY}) confirm the presence of an extended dusty envelope. The large-scale geometry of the star has already been imaged in the thermal infrared with TIMMI in the $N$-band \citep{Lagadec} and with NACO in the $L$-band \citep{Lagadec_L_2007apn4.confE..16L}. These observations, together with the VISIR coronagraphic image reveal the presence of an extended  emission region (with a radius up to 2~arcsec). 
However, the limited angular resolution capabilities of the large-scale images do not allow the authors to resolve the innermost part of the structure. As a consequence, these observations bring information only on the extended structure and can be seen as fully complementary to the image provided by MATISSE. 

One question that arises is the overall geometry of the CSE. If infrared observations favour an extended CSE (elliptical in the MATISSE field of view but nearly circular at larger scales), other observations at different wavelengths (e.g. \citealt{SahaiALMA}) favour an overall geometry consisting of a combination of a bipolar high-velocity wind and an equatorial dense disc fed by a slow wind. This scenario is compatible with the different CO-outflows found  by \cite{Knapp_1997A&A...326..318K}, and with the high-velocity ejection \citep{Sahai_HST}. More recently, the dense disc was revealed by asymmetric emission in near-infrared scattered light \citep{Sahai_2019IAUS..343..495S}. 
These two possible geometries (dense equatorial disc and/or high-velocity bipolar outflow) can neither be confirmed nor dismissed by the thermal infrared imaging (both MATISSE and previous observations).

Indeed, the high-velocity gaseous bipolar outflow could not be detected in the images because the low spectral resolution mode ($R=30$) does not allow us to characterise the velocity structure of the emission, contrary to what was done with Brackett lines in $\eta$ Car where a spectral-channel width of 20 - 40~$\rm km\,s^{-1}$ was achieved for Br$\gamma$ in the $K$-band, and of 170~$\rm km\,s^{-1}$
 for Br$\alpha$ in the $L$-band (\citealt{etaCar_AMBER_2016A&A...594A.106W}, \citealt{eta_Car_GRAVITY_2018A&A...618A.125G}, \citealt{eta_Car_MATISSE_2021A&A...652A.140W}). 
   
Regarding the presence of an equatorial density enhancement, given the low inclination of the system, such disc-like configuration emitting in the infrared would lead to a large emission region slightly elongated on a 2D-image, making it indistinguishable from an elongated dusty CSE. 
If a disc is responsible for the extended emission instead of a CSE, it should be circumbinary. Such discs are extensively found in post-AGB binaries \citep{Kluska_post_AGB_disc_2022A&A...658A..36K}. 
As no signature of the disc is found in the MATISSE images, its inner radius should be located outside the field of view of the instrument.



\section{Summary and conclusions}
\label{sect:conclusion}
In this paper, we present the reconstructed images of V~Hya in the thermal infrared obtained with the MATISSE instrument.
We first modelled the spectral energy distribution using a composite model combining a hydrostatic COMARCS spectrum with DUSTY. It reveals that the star is surrounded by 
an extended dusty CSE, made of amorphous carbon and silicon carbide, that dominates the emission in the thermal infrared. 

We performed a preliminary model-fitting of the visibility measurements. Then, we performed image reconstruction using the visibility and closure phase signals. The images of V~Hya obtained from MATISSE interferometric measurements were processed using SQUEEZE, MIRA and IRBis. In the $L$-band we presented the stellar photosphere in a molecular absorption band of $\rm C_2H_2$-HCN and in the pseudo-continuum. We showed that both images display a non-spherically symmetric structure and explained it by photospheric material ejection. 
In the $N$-band we revealed the presence of a North-East 20~mas extension to the stellar photosphere. We concluded that the extension arises from a dust enhancement whose position roughly matches the prediction for the companion position. We compared our MATISSE image with archive data from MIDI and observed changes in the interferometric data between the two epochs, compatible with the orbital motion. Hence, the presence of a dusty clump around a companion is favoured compared to a blob ejected in a random direction by a single star.
Finally, we 
compared the theoretical radial profile from the SED modelling with the observed composite radial profile constrained at the small scales by the MATISSE $N$-band image and at large scales by the images obtained with VISIR. We confirmed the presence of the extended dusty CSE around the central source.
The high-velocity gaseous jet is not detected in the image and would require higher spectral resolution to resolve its origin and its kinematic structure. 

Future observations at different epochs would be of particular interest to retrieve temporal information about the dusty wind as a function of the orbital motion, and would provide strong constraints on the dust distribution around the binary components. 
Eventually, multi-dimensional radiative transfer reproducing the circumstellar and circumbinary environment of the V Hya system should be foreseen to connect the binary interaction to the large-scale morphology. 

\begin{acknowledgements}
     L.P. acknowledges an ESO studentship, and is FNRS research fellow. A.J. is supported by the {\em Fonds National de la Recherche Scientifique} (F.R.S.-FNRS) under PDR grant T.0115.23.
    S.H. acknowledges funding from the European Research Council (ERC) under the European Union’s Horizon 2020 research and innovation programme (Grant agreement No. 883867, project EXWINGS) and the Swedish Research Council (Vetenskapsradet, grant number 2019-04059).

    F.L. received funding from the Hungarian NKFIH OTKA project no. K-132406.

     M.M. acknowledges funding from the Programme Paris Region fellowship supported by the R\'egion Ile-de-France. This project has received funding from the European Union’s Horizon 2020 research and innovation program under the Marie Skłodowska-Curie Grant agreement No. 945298.
     
    K.O. acknowleges the support of the Agencia Nacional de Investigaci\'on Cient\'\i fica y Desarrollo (ANID) through the FONDECYT Regular grant 1210652.
This research has made use of the Jean-Marie Mariotti Center \texttt{Aspro} (\url{http://www.jmmc.fr/aspro}) and \texttt{OIFits Explorer} (\url{http://www.jmmc.fr/oifitsexplorer}) services.
     Use was made of the Simbad database, operated at the CDS, Strasbourg, France, and of NASA’s Astrophysics Data System Bibliographic Services. This research made use of Numpy, Matplotlib, SciPy and Astropy, a community-developed core Python package for Astronomy \citep{Astropy_2018AJ....156..123A}. 
     We acknowledge with thanks the variable star observations from the AAVSO International Database contributed by observers worldwide and used in this research.
     
\end{acknowledgements}

\bibliographystyle{aa} 
\bibliography{references} 

\begin{thebibliography}{81}
\expandafter\ifx\csname natexlab\endcsname\relax\def\natexlab#1{#1}\fi

\bibitem[{{Andersen} {et~al.}(1999){Andersen}, {Loidl}, \&
  {H{\"o}fner}}]{ref_QA_1999A&A...349..243A}
{Andersen}, A.~C., {Loidl}, R., \& {H{\"o}fner}, S. 1999, \aap, 349, 243

\bibitem[{{Andriantsaralaza} {et~al.}(2022){Andriantsaralaza}, {Ramstedt},
  {Vlemmings}, \& {De Beck}}]{Gaia_AGB_priors_2022A&A...667A..74A}
{Andriantsaralaza}, M., {Ramstedt}, S., {Vlemmings}, W.~H.~T., \& {De Beck}, E.
  2022, \aap, 667, A74

\bibitem[{{Aringer} {et~al.}(2009){Aringer}, {Girardi}, {Nowotny}, {Marigo}, \&
  {Lederer}}]{COMARCS}
{Aringer}, B., {Girardi}, L., {Nowotny}, W., {Marigo}, P., \& {Lederer}, M.~T.
  2009, \aap, 503, 913

\bibitem[{{Astropy Collaboration} {et~al.}(2018){Astropy Collaboration},
  {Price-Whelan}, {Sip{\H{o}}cz}, {G{\"u}nther}, {Lim}, {Crawford}, {Conseil},
  {Shupe}, {Craig}, {Dencheva}, {Ginsburg}, {VanderPlas}, {Bradley},
  {P{\'e}rez-Su{\'a}rez}, {de Val-Borro}, {Aldcroft}, {Cruz}, {Robitaille},
  {Tollerud}, {Ardelean}, {Babej}, {Bach}, {Bachetti}, {Bakanov}, {Bamford},
  {Barentsen}, {Barmby}, {Baumbach}, {Berry}, {Biscani}, {Boquien}, {Bostroem},
  {Bouma}, {Brammer}, {Bray}, {Breytenbach}, {Buddelmeijer}, {Burke},
  {Calderone}, {Cano Rodr{\'\i}guez}, {Cara}, {Cardoso}, {Cheedella}, {Copin},
  {Corrales}, {Crichton}, {D'Avella}, {Deil}, {Depagne}, {Dietrich}, {Donath},
  {Droettboom}, {Earl}, {Erben}, {Fabbro}, {Ferreira}, {Finethy}, {Fox},
  {Garrison}, {Gibbons}, {Goldstein}, {Gommers}, {Greco}, {Greenfield},
  {Groener}, {Grollier}, {Hagen}, {Hirst}, {Homeier}, {Horton}, {Hosseinzadeh},
  {Hu}, {Hunkeler}, {Ivezi{\'c}}, {Jain}, {Jenness}, {Kanarek}, {Kendrew},
  {Kern}, {Kerzendorf}, {Khvalko}, {King}, {Kirkby}, {Kulkarni}, {Kumar},
  {Lee}, {Lenz}, {Littlefair}, {Ma}, {Macleod}, {Mastropietro}, {McCully},
  {Montagnac}, {Morris}, {Mueller}, {Mumford}, {Muna}, {Murphy}, {Nelson},
  {Nguyen}, {Ninan}, {N{\"o}the}, {Ogaz}, {Oh}, {Parejko}, {Parley}, {Pascual},
  {Patil}, {Patil}, {Plunkett}, {Prochaska}, {Rastogi}, {Reddy Janga},
  {Sabater}, {Sakurikar}, {Seifert}, {Sherbert}, {Sherwood-Taylor}, {Shih},
  {Sick}, {Silbiger}, {Singanamalla}, {Singer}, {Sladen}, {Sooley},
  {Sornarajah}, {Streicher}, {Teuben}, {Thomas}, {Tremblay}, {Turner},
  {Terr{\'o}n}, {van Kerkwijk}, {de la Vega}, {Watkins}, {Weaver}, {Whitmore},
  {Woillez}, {Zabalza}, \& {Astropy
  Contributors}}]{Astropy_2018AJ....156..123A}
{Astropy Collaboration}, {Price-Whelan}, A.~M., {Sip{\H{o}}cz}, B.~M., {et~al.}
  2018, \aj, 156, 123

\bibitem[{{Bailer-Jones}(2015)}]{Bailer-Jones2015}
{Bailer-Jones}, C. A.~L. 2015, \pasp, 127, 994

\bibitem[{{Barnbaum} {et~al.}(1995){Barnbaum}, {Morris}, \&
  {Kahane}}]{Barnbaum_1995ApJ...450..862B}
{Barnbaum}, C., {Morris}, M., \& {Kahane}, C. 1995, \apj, 450, 862

\bibitem[{{Baron} {et~al.}(2010){Baron}, {Monnier}, \& {Kloppenborg}}]{SQUEEZE}
{Baron}, F., {Monnier}, J.~D., \& {Kloppenborg}, B. 2010, in Society of
  Photo-Optical Instrumentation Engineers (SPIE) Conference Series, Vol. 7734,
  Optical and Infrared Interferometry II, ed. W.~C. {Danchi}, F.~{Delplancke},
  \& J.~K. {Rajagopal}, 77342I

\bibitem[{{Bergeat} {et~al.}(1998){Bergeat}, {Knapik}, \&
  {Rutily}}]{Bergeat_1998A&A...332L..53B}
{Bergeat}, J., {Knapik}, A., \& {Rutily}, B. 1998, \aap, 332, L53

\bibitem[{{Bourges} {et~al.}(2017){Bourges}, {Mella}, {Lafrasse}, {Duvert},
  {Chelli}, {Le Bouquin}, {Delfosse}, \& {Chesneau}}]{JMMC_vizier_II/346}
{Bourges}, L., {Mella}, G., {Lafrasse}, S., {et~al.} 2017, JMMC Stellar
  Diameters Catalogue - JSDC. Version 2

\bibitem[{{Chen} {et~al.}(2020){Chen}, {Ivanova}, \&
  {Carroll-Nellenback}}]{Chen2020}
{Chen}, Z., {Ivanova}, N., \& {Carroll-Nellenback}, J. 2020, \apj, 892, 110

\bibitem[{{Chiavassa} {et~al.}(2011){Chiavassa}, {Pasquato}, {Jorissen},
  {Sacuto}, {Babusiaux}, {Freytag}, {Ludwig}, {Cruzal{\`e}bes}, {Rabbia},
  {Spang}, \& {Chesneau}}]{2011A&A...528A.120C}
{Chiavassa}, A., {Pasquato}, E., {Jorissen}, A., {et~al.} 2011, \aap, 528, A120

\bibitem[{{Cohen} {et~al.}(2003){Cohen}, {Megeath}, {Hammersley},
  {Mart{\'\i}n-Luis}, \& {Stauffer}}]{2003AJ....125.2645C}
{Cohen}, M., {Megeath}, S.~T., {Hammersley}, P.~L., {Mart{\'\i}n-Luis}, F., \&
  {Stauffer}, J. 2003, \aj, 125, 2645

\bibitem[{{Cruzal{\`e}bes} {et~al.}(2019){Cruzal{\`e}bes}, {Petrov},
  {Robbe-Dubois}, {Varga}, {Burtscher}, {Allouche}, {Berio}, {Hofmann}, {Hron},
  {Jaffe}, {Lagarde}, {Lopez}, {Matter}, {Meilland}, {Meisenheimer}, {Millour},
  \& {Schertl}}]{vizier:II/361}
{Cruzal{\`e}bes}, P., {Petrov}, R., {Robbe-Dubois}, S., {et~al.} 2019, MDFC
  Version 10

\bibitem[{{Danchi} {et~al.}(1994){Danchi}, {Bester}, {Degiacomi}, {Greenhill},
  \& {Townes}}]{1994AJ....107.1469D}
{Danchi}, W.~C., {Bester}, M., {Degiacomi}, C.~G., {Greenhill}, L.~J., \&
  {Townes}, C.~H. 1994, \aj, 107, 1469

\bibitem[{{Decin} {et~al.}(2020){Decin}, {Montarg{\`e}s}, {Richards},
  {Gottlieb}, {Homan}, {McDonald}, {El Mellah}, {Danilovich}, {Wallstr{\"o}m},
  {Zijlstra}, {Baudry}, {Bolte}, {Cannon}, {De Beck}, {De Ceuster}, {de Koter},
  {De Ridder}, {Etoka}, {Gobrecht}, {Gray}, {Herpin}, {Jeste}, {Lagadec},
  {Kervella}, {Khouri}, {Menten}, {Millar}, {M{\"u}ller}, {Plane}, {Sahai},
  {Sana}, {Van de Sande}, {Waters}, {Wong}, \& {Yates}}]{2020Sci...369.1497D}
{Decin}, L., {Montarg{\`e}s}, M., {Richards}, A.~M.~S., {et~al.} 2020, Science,
  369, 1497

\bibitem[{Delacroix {et~al.}(2012)Delacroix, Absil, Mawet, Hanot, Karlsson,
  Forsberg, Pantin, Surdej, \& Habraken}]{2012SPIE.8446E..8KD}
Delacroix, C., Absil, O., Mawet, D., {et~al.} 2012, in Ground-based and
  Airborne Instrumentation for Astronomy IV. Proceedings of the SPIE, Univ. de
  Li{\`e}ge (Belgium)

\bibitem[{{Doan} {et~al.}(2020){Doan}, {Ramstedt}, {Vlemmings}, {Mohamed},
  {H{\"o}fner}, {De Beck}, {Kerschbaum}, {Lindqvist}, {Maercker}, {Paladini},
  \& {Wittkowski}}]{WAql_2020A&A...633A..13D}
{Doan}, L., {Ramstedt}, S., {Vlemmings}, W.~H.~T., {et~al.} 2020, \aap, 633,
  A13

\bibitem[{{Drevon} {et~al.}(2022{\natexlab{a}}){Drevon}, {Millour},
  {Cruzal{\`e}bes}, {Paladini}, {Hron}, {Meilland}, {Allouche}, {Hofmann},
  {Lagarde}, {Lopez}, {Matter}, {Petrov}, {Robbe-Dubois}, {Schertl},
  {Scicluna}, {Wittkowski}, {Zins}, {{\'A}brah{\'a}m}, {Antonelli}, {Beckmann},
  {Berio}, {Bettonvil}, {Glindemann}, {Graser}, {Heininger}, {Henning},
  {Isbell}, {Jaffe}, {Labadie}, {Leinert}, {Lehmitz}, {Morel}, {Meisenheimer},
  {Soulain}, {Varga}, {Weigelt}, {Woillez}, {Augereau}, {van Boekel},
  {Burtscher}, {Danchi}, {Dominik}, {G{\'a}mez Rosas}, {Hocd{\'e}},
  {Hogerheijde}, {Klarmann}, {Kokoulina}, {Leftley}, {Stee}, {Vakili},
  {Waters}, {Wolf}, \& {Yoffe}}]{2022A&A...665A..32D}
{Drevon}, J., {Millour}, F., {Cruzal{\`e}bes}, P., {et~al.} 2022{\natexlab{a}},
  \aap, 665, A32

\bibitem[{{Drevon} {et~al.}(2022{\natexlab{b}}){Drevon}, {Millour},
  {Cruzal{\`e}bes}, {Scicluna}, \& {Paladini}}]{RHAPSODY}
{Drevon}, J., {Millour}, F., {Cruzal{\`e}bes}, P., {Scicluna}, P., \&
  {Paladini}, C. 2022{\natexlab{b}}, in Society of Photo-Optical
  Instrumentation Engineers (SPIE) Conference Series, Vol. 12183, Optical and
  Infrared Interferometry and Imaging VIII, ed. A.~{M{\'e}rand}, S.~{Sallum},
  \& J.~{Sanchez-Bermudez}, 121831O

\bibitem[{{Ferreira} {et~al.}(2007){Ferreira}, {Dougados}, \&
  {Whelan}}]{Ferreira_2007LNP...723.....F}
{Ferreira}, J., {Dougados}, C., \& {Whelan}, E. 2007, {Jets from Young Stars I:
  Models and Constraints}, Vol. 723

\bibitem[{{Freytag} \& {H{\"o}fner}(2023)}]{Freytag2023}
{Freytag}, B. \& {H{\"o}fner}, S. 2023, \aap, 669, A155

\bibitem[{{Gaia Collaboration} {et~al.}(2021){Gaia Collaboration}, {Brown},
  {Vallenari}, {Prusti}, {de Bruijne}, {Babusiaux}, {Biermann}, {Creevey},
  {Evans}, {Eyer}, {Hutton}, {Jansen}, {Jordi}, {Klioner}, {Lammers},
  {Lindegren}, {Luri}, {Mignard}, {Panem}, {Pourbaix}, {Randich}, {Sartoretti},
  {Soubiran}, {Walton}, {Arenou}, {Bailer-Jones}, {Bastian}, {Cropper},
  {Drimmel}, {Katz}, {Lattanzi}, {van Leeuwen}, {Bakker}, {Cacciari},
  {Casta{\~n}eda}, {De Angeli}, {Ducourant}, {Fabricius}, {Fouesneau},
  {Fr{\'e}mat}, {Guerra}, {Guerrier}, {Guiraud}, {Jean-Antoine Piccolo},
  {Masana}, {Messineo}, {Mowlavi}, {Nicolas}, {Nienartowicz}, {Pailler},
  {Panuzzo}, {Riclet}, {Roux}, {Seabroke}, {Sordo}, {Tanga}, {Th{\'e}venin},
  {Gracia-Abril}, {Portell}, {Teyssier}, {Altmann}, {Andrae}, {Bellas-Velidis},
  {Benson}, {Berthier}, {Blomme}, {Brugaletta}, {Burgess}, {Busso}, {Carry},
  {Cellino}, {Cheek}, {Clementini}, {Damerdji}, {Davidson}, {Delchambre},
  {Dell'Oro}, {Fern{\'a}ndez-Hern{\'a}ndez}, {Galluccio}, {Garc{\'\i}a-Lario},
  {Garcia-Reinaldos}, {Gonz{\'a}lez-N{\'u}{\~n}ez}, {Gosset}, {Haigron},
  {Halbwachs}, {Hambly}, {Harrison}, {Hatzidimitriou}, {Heiter},
  {Hern{\'a}ndez}, {Hestroffer}, {Hodgkin}, {Holl}, {Jan{\ss}en}, {Jevardat de
  Fombelle}, {Jordan}, {Krone-Martins}, {Lanzafame}, {L{\"o}ffler}, {Lorca},
  {Manteiga}, {Marchal}, {Marrese}, {Moitinho}, {Mora}, {Muinonen}, {Osborne},
  {Pancino}, {Pauwels}, {Petit}, {Recio-Blanco}, {Richards}, {Riello},
  {Rimoldini}, {Robin}, {Roegiers}, {Rybizki}, {Sarro}, {Siopis}, {Smith},
  {Sozzetti}, {Ulla}, {Utrilla}, {van Leeuwen}, {van Reeven}, {Abbas}, {Abreu
  Aramburu}, {Accart}, {Aerts}, {Aguado}, {Ajaj}, {Altavilla}, {{\'A}lvarez},
  {{\'A}lvarez Cid-Fuentes}, {Alves}, {Anderson}, {Anglada Varela}, {Antoja},
  {Audard}, {Baines}, {Baker}, {Balaguer-N{\'u}{\~n}ez}, {Balbinot}, {Balog},
  {Barache}, {Barbato}, {Barros}, {Barstow}, {Bartolom{\'e}}, {Bassilana},
  {Bauchet}, {Baudesson-Stella}, {Becciani}, {Bellazzini}, {Bernet}, {Bertone},
  {Bianchi}, {Blanco-Cuaresma}, {Boch}, {Bombrun}, {Bossini}, {Bouquillon},
  {Bragaglia}, {Bramante}, {Breedt}, {Bressan}, {Brouillet}, {Bucciarelli},
  {Burlacu}, {Busonero}, {Butkevich}, {Buzzi}, {Caffau}, {Cancelliere},
  {C{\'a}novas}, {Cantat-Gaudin}, {Carballo}, {Carlucci}, {Carnerero},
  {Carrasco}, {Casamiquela}, {Castellani}, {Castro-Ginard}, {Castro Sampol},
  {Chaoul}, {Charlot}, {Chemin}, {Chiavassa}, {Cioni}, {Comoretto}, {Cooper},
  {Cornez}, {Cowell}, {Crifo}, {Crosta}, {Crowley}, {Dafonte}, {Dapergolas},
  {David}, {David}, {de Laverny}, {De Luise}, {De March}, {De Ridder}, {de
  Souza}, {de Teodoro}, {de Torres}, {del Peloso}, {del Pozo}, {Delbo},
  {Delgado}, {Delgado}, {Delisle}, {Di Matteo}, {Diakite}, {Diener},
  {Distefano}, {Dolding}, {Eappachen}, {Edvardsson}, {Enke}, {Esquej}, {Fabre},
  {Fabrizio}, {Faigler}, {Fedorets}, {Fernique}, {Fienga}, {Figueras},
  {Fouron}, {Fragkoudi}, {Fraile}, {Franke}, {Gai}, {Garabato},
  {Garcia-Gutierrez}, {Garc{\'\i}a-Torres}, {Garofalo}, {Gavras}, {Gerlach},
  {Geyer}, {Giacobbe}, {Gilmore}, {Girona}, {Giuffrida}, {Gomel}, {Gomez},
  {Gonzalez-Santamaria}, {Gonz{\'a}lez-Vidal}, {Granvik},
  {Guti{\'e}rrez-S{\'a}nchez}, {Guy}, {Hauser}, {Haywood}, {Helmi}, {Hidalgo},
  {Hilger}, {H{\l}adczuk}, {Hobbs}, {Holland}, {Huckle}, {Jasniewicz},
  {Jonker}, {Juaristi Campillo}, {Julbe}, {Karbevska}, {Kervella}, {Khanna},
  {Kochoska}, {Kontizas}, {Kordopatis}, {Korn}, {Kostrzewa-Rutkowska},
  {Kruszy{\'n}ska}, {Lambert}, {Lanza}, {Lasne}, {Le Campion}, {Le Fustec},
  {Lebreton}, {Lebzelter}, {Leccia}, {Leclerc}, {Lecoeur-Taibi}, {Liao},
  {Licata}, {Lindstr{\o}m}, {Lister}, {Livanou}, {Lobel}, {Madrero Pardo},
  {Managau}, {Mann}, {Marchant}, {Marconi}, {Marcos Santos}, {Marinoni},
  {Marocco}, {Marshall}, {Martin Polo}, {Mart{\'\i}n-Fleitas}, {Masip},
  {Massari}, {Mastrobuono-Battisti}, {Mazeh}, {McMillan}, {Messina},
  {Michalik}, {Millar}, {Mints}, {Molina}, {Molinaro}, {Moln{\'a}r},
  {Montegriffo}, {Mor}, {Morbidelli}, {Morel}, {Morris}, {Mulone}, {Munoz},
  {Muraveva}, {Murphy}, {Musella}, {Noval}, {Ord{\'e}novic}, {Orr{\`u}},
  {Osinde}, {Pagani}, {Pagano}, {Palaversa}, {Palicio}, {Panahi}, {Pawlak},
  {Pe{\~n}alosa Esteller}, {Penttil{\"a}}, {Piersimoni}, {Pineau}, {Plachy},
  {Plum}, {Poggio}, {Poretti}, {Poujoulet}, {Pr{\v{s}}a}, {Pulone}, {Racero},
  {Ragaini}, {Rainer}, {Raiteri}, {Rambaux}, {Ramos}, {Ramos-Lerate}, {Re
  Fiorentin}, {Regibo}, {Reyl{\'e}}, {Ripepi}, {Riva}, {Rixon}, {Robichon},
  {Robin}, {Roelens}, {Rohrbasser}, {Romero-G{\'o}mez}, {Rowell}, {Royer},
  {Rybicki}, {Sadowski}, {Sagrist{\`a} Sell{\'e}s}, {Sahlmann}, {Salgado},
  {Salguero}, {Samaras}, {Sanchez Gimenez}, {Sanna}, {Santove{\~n}a},
  {Sarasso}, {Schultheis}, {Sciacca}, {Segol}, {Segovia}, {S{\'e}gransan},
  {Semeux}, {Shahaf}, {Siddiqui}, {Siebert}, {Siltala}, {Slezak}, {Smart},
  {Solano}, {Solitro}, {Souami}, {Souchay}, {Spagna}, {Spoto}, {Steele},
  {Steidelm{\"u}ller}, {Stephenson}, {S{\"u}veges}, {Szabados}, {Szegedi-Elek},
  {Taris}, {Tauran}, {Taylor}, {Teixeira}, {Thuillot}, {Tonello}, {Torra},
  {Torra}, {Turon}, {Unger}, {Vaillant}, {van Dillen}, {Vanel}, {Vecchiato},
  {Viala}, {Vicente}, {Voutsinas}, {Weiler}, {Wevers}, {Wyrzykowski}, {Yoldas},
  {Yvard}, {Zhao}, {Zorec}, {Zucker}, {Zurbach}, \&
  {Zwitter}}]{GDR3_2022arXiv220800211G}
{Gaia Collaboration}, {Brown}, A.~G.~A., {Vallenari}, A., {et~al.} 2021, \aap,
  649, A1

\bibitem[{{G{\'a}mez Rosas} {et~al.}(2022){G{\'a}mez Rosas}, {Isbell}, {Jaffe},
  {Petrov}, {Leftley}, {Hofmann}, {Millour}, {Burtscher}, {Meisenheimer},
  {Meilland}, {Waters}, {Lopez}, {Lagarde}, {Weigelt}, {Berio}, {Allouche},
  {Robbe-Dubois}, {Cruzal{\`e}bes}, {Bettonvil}, {Henning}, {Augereau},
  {Antonelli}, {Beckmann}, {van Boekel}, {Bendjoya}, {Danchi}, {Dominik},
  {Drevon}, {Gallimore}, {Graser}, {Heininger}, {Hocd{\'e}}, {Hogerheijde},
  {Hron}, {Impellizzeri}, {Klarmann}, {Kokoulina}, {Labadie}, {Lehmitz},
  {Matter}, {Paladini}, {Pantin}, {Pott}, {Schertl}, {Soulain}, {Stee},
  {Tristram}, {Varga}, {Woillez}, {Wolf}, {Yoffe}, \&
  {Zins}}]{AGN_2022Natur.602..403G}
{G{\'a}mez Rosas}, V., {Isbell}, J.~W., {Jaffe}, W., {et~al.} 2022, \nat, 602,
  403

\bibitem[{{Gordon} {et~al.}(2009){Gordon}, {Cartledge}, \&
  {Clayton}}]{MW_exctinction_2009ApJ...705.1320G}
{Gordon}, K.~D., {Cartledge}, S., \& {Clayton}, G.~C. 2009, \apj, 705, 1320

\bibitem[{{GRAVITY Collaboration} {et~al.}(2018){GRAVITY Collaboration},
  {Sanchez-Bermudez}, {Weigelt}, {Bestenlehner}, {Kervella}, {Brandner},
  {Henning}, {M{\"u}ller}, {Perrin}, {Pott}, {Sch{\"o}ller}, {van Boekel},
  {Abuter}, {Accardo}, {Amorim}, {Anugu}, {{\'A}vila}, {Benisty}, {Berger},
  {Blind}, {Bonnet}, {Bourget}, {Brast}, {Buron}, {Cantalloube}, {Caratti O
  Garatti}, {Cassaing}, {Chapron}, {Choquet}, {Cl{\'e}net}, {Collin},
  {Coud{\'e} Du Foresto}, {de Wit}, {de Zeeuw}, {Deen},
  {Delplancke-Str{\"o}bele}, {Dembet}, {Derie}, {Dexter}, {Duvert}, {Ebert},
  {Eckart}, {Eisenhauer}, {Esselborn}, {F{\'e}dou}, {Garcia}, {Garcia Dabo},
  {Garcia Lopez}, {Gao}, {Gendron}, {Genzel}, {Gillessen}, {Haubois}, {Haug},
  {Haussmann}, {Hippler}, {Horrobin}, {Huber}, {Hubert}, {Hubin}, {Hummel},
  {Jakob}, {Jochum}, {Jocou}, {Karl}, {Kaufer}, {Kellner}, {Kendrew}, {Kern},
  {Kiekebusch}, {Klein}, {Kolb}, {Kulas}, {Lacour}, {Lapeyr{\`e}re},
  {Lazareff}, {Le Bouquin}, {L{\'e}na}, {Lenzen}, {L{\'e}v{\^e}que}, {Lippa},
  {Magnard}, {Mehrgan}, {Mellein}, {M{\'e}rand}, {Moreno-Ventas}, {Moulin},
  {M{\"u}ller}, {M{\"u}ller}, {Neumann}, {Oberti}, {Ott}, {Pallanca},
  {Panduro}, {Pasquini}, {Paumard}, {Percheron}, {Perraut}, {Petrucci},
  {Pfl{\"u}ger}, {Pfuhl}, {Duc}, {Plewa}, {Popovic}, {Rabien}, {Ramirez},
  {Ramos}, {Rau}, {Riquelme}, {Rodr{\'\i}guez-Coira}, {Rohloff}, {Rosales},
  {Rousset}, {Scheithauer}, {Schuhler}, {Spyromilio}, {Straub}, {Straubmeier},
  {Sturm}, {Suarez}, {Tristram}, {Ventura}, {Vincent}, {Waisberg}, {Wank},
  {Widmann}, {Wieprecht}, {Wiest}, {Wiezorrek}, {Wittkowski}, {Woillez},
  {Wolff}, {Yazici}, {Ziegler}, \&
  {Zins}}]{eta_Car_GRAVITY_2018A&A...618A.125G}
{GRAVITY Collaboration}, {Sanchez-Bermudez}, J., {Weigelt}, G., {et~al.} 2018,
  \aap, 618, A125

\bibitem[{{Hanner}(1988)}]{refSiC_1988ioch.rept.....H}
{Hanner}, M.~S. 1988, {Infrared observations of comets Halley and Wilson and
  properties of the grains : summary of workshop sponsored by the National
  Aeronautics and Space Administration, Washington, D.C. and held at Cornell
  University, Ithaca, New York, August 10-12, 1987}, NASA Conference
  Publication, Summary of a Workshop, held at Cornell University, Ithaca, New
  York, August 10-12, 1987, Washington: NASA, 1988, edited by Hanner, Martha S.

\bibitem[{{Hirano} {et~al.}(2004){Hirano}, {Shinnaga}, {Dinh-V-Trung}, {Fong},
  {Keto}, {Patel}, {Qi}, {Young}, {Zhang}, \&
  {Zhao}}]{referee_suggestion_Hirano_CO_radio_wind_2004ApJ...616L..43H}
{Hirano}, N., {Shinnaga}, H., {Dinh-V-Trung}, {et~al.} 2004, \apjl, 616, L43

\bibitem[{{Hofmann} {et~al.}(2014){Hofmann}, {Weigelt}, \& {Schertl}}]{IRBis}
{Hofmann}, K.~H., {Weigelt}, G., \& {Schertl}, D. 2014, \aap, 565, A48

\bibitem[{{H{\"o}fner} \& {Freytag}(2019)}]{Suzanne_ref_2019A&A...623A.158H}
{H{\"o}fner}, S. \& {Freytag}, B. 2019, \aap, 623, A158

\bibitem[{{H{\"o}fner} \&
  {Olofsson}(2018)}]{Hofner_Olofsson_2018A&ARv..26....1H}
{H{\"o}fner}, S. \& {Olofsson}, H. 2018, \aapr, 26, 1

\bibitem[{{Ivezic} \& {Elitzur}(1997)}]{DUSTY}
{Ivezic}, Z. \& {Elitzur}, M. 1997, \mnras, 287, 799

\bibitem[{{Jones} \& {Boffin}(2017)}]{PN_claire_ref_2017NatAs...1E.117J}
{Jones}, D. \& {Boffin}, H. M.~J. 2017, Nature Astronomy, 1, 0117

\bibitem[{{J{\o}rgensen} {et~al.}(2000){J{\o}rgensen}, {Hron}, \&
  {Loidl}}]{2000A&A...356..253J}
{J{\o}rgensen}, U.~G., {Hron}, J., \& {Loidl}, R. 2000, \aap, 356, 253

\bibitem[{{Kahane} {et~al.}(1996){Kahane}, {Audinos}, {Barnbaum}, \&
  {Morris}}]{Kahane_1996A&A...314..871K}
{Kahane}, C., {Audinos}, P., {Barnbaum}, C., \& {Morris}, M. 1996, \aap, 314,
  871

\bibitem[{{Karovska} {et~al.}(2004){Karovska}, {Wood}, {Marengo}, {Raymond},
  {Hack}, \& {Guinan}}]{Mira_AB_2004RMxAC..20...92K}
{Karovska}, M., {Wood}, B., {Marengo}, M., {et~al.} 2004, in Revista Mexicana
  de Astronomia y Astrofisica Conference Series, ed. G.~{Tovmassian} \&
  E.~{Sion}, Vol.~20, 92--94

\bibitem[{{Kervella} {et~al.}(2016){Kervella}, {Homan}, {Richards}, {Decin},
  {McDonald}, {Montarg{\`e}s}, \& {Ohnaka}}]{Kervella_2016A&A...596A..92K}
{Kervella}, P., {Homan}, W., {Richards}, A.~M.~S., {et~al.} 2016, \aap, 596,
  A92

\bibitem[{{Kloppenborg}(2023)}]{AAVSO_ref}
{Kloppenborg}, B.~K. 2023, Observations from the AAVSO International Database,
  \url{https://www.aavso.org}

\bibitem[{{Klotz} {et~al.}(2012){Klotz}, {Sacuto}, {Paladini}, {Hron}, \&
  {Wachter}}]{GEM-FIND}
{Klotz}, D., {Sacuto}, S., {Paladini}, C., {Hron}, J., \& {Wachter}, G. 2012,
  in Society of Photo-Optical Instrumentation Engineers (SPIE) Conference
  Series, Vol. 8445, Optical and Infrared Interferometry III, ed.
  F.~{Delplancke}, J.~K. {Rajagopal}, \& F.~{Malbet}, 84451A

\bibitem[{{Kluska} {et~al.}(2022){Kluska}, {Van Winckel}, {Copp{\'e}e},
  {Oomen}, {Dsilva}, {Kamath}, {Bujarrabal}, \&
  {Min}}]{Kluska_post_AGB_disc_2022A&A...658A..36K}
{Kluska}, J., {Van Winckel}, H., {Copp{\'e}e}, Q., {et~al.} 2022, \aap, 658,
  A36

\bibitem[{{Knapp} {et~al.}(1999){Knapp}, {Dobrovolsky}, {Ivezi{\'c}}, {Young},
  {Crosas}, {Mattei}, \& {Rupen}}]{KNAPP_1999A&A...351...97K}
{Knapp}, G.~R., {Dobrovolsky}, S.~I., {Ivezi{\'c}}, Z., {et~al.} 1999, \aap,
  351, 97

\bibitem[{{Knapp} {et~al.}(1997){Knapp}, {Jorissen}, \&
  {Young}}]{Knapp_1997A&A...326..318K}
{Knapp}, G.~R., {Jorissen}, A., \& {Young}, K. 1997, \aap, 326, 318

\bibitem[{{Lagadec} {et~al.}(2007){Lagadec}, {Chesneau}, {Zijlstra},
  {Matsuura}, \& {M{\'e}karnia}}]{Lagadec_L_2007apn4.confE..16L}
{Lagadec}, E., {Chesneau}, O., {Zijlstra}, A.~A., {Matsuura}, M., \&
  {M{\'e}karnia}, D. 2007, in Asymmetrical Planetary Nebulae IV, 16

\bibitem[{{Lagadec} {et~al.}(2005){Lagadec}, {M{\'e}karnia}, {de Freitas
  Pacheco}, \& {Dougados}}]{Lagadec}
{Lagadec}, E., {M{\'e}karnia}, D., {de Freitas Pacheco}, J.~A., \& {Dougados},
  C. 2005, \aap, 433, 553

\bibitem[{{Lagage} {et~al.}(2004){Lagage}, {Pel}, {Authier}, {Belorgey},
  {Claret}, {Doucet}, {Dubreuil}, {Durand}, {Elswijk}, {Girardot}, {K{\"a}ufl},
  {Kroes}, {Lortholary}, {Lussignol}, {Marchesi}, {Pantin}, {Peletier},
  {Pirard}, {Pragt}, {Rio}, {Schoenmaker}, {Siebenmorgen}, {Silber}, {Smette},
  {Sterzik}, \& {Veyssiere}}]{VISIR_2004Msngr.117...12L}
{Lagage}, P.~O., {Pel}, J.~W., {Authier}, M., {et~al.} 2004, The Messenger,
  117, 12

\bibitem[{{Lallement} {et~al.}(2019){Lallement}, {Babusiaux}, {Vergely},
  {Katz}, {Arenou}, {Valette}, {Hottier}, \&
  {Capitanio}}]{Lallement_2019A&A...625A.135L}
{Lallement}, R., {Babusiaux}, C., {Vergely}, J.~L., {et~al.} 2019, \aap, 625,
  A135

\bibitem[{{Lambert} {et~al.}(1986){Lambert}, {Gustafsson}, {Eriksson}, \&
  {Hinkle}}]{1986ApJS...62..373L}
{Lambert}, D.~L., {Gustafsson}, B., {Eriksson}, K., \& {Hinkle}, K.~H. 1986,
  \apjs, 62, 373

\bibitem[{{Lederer} \& {Aringer}(2009)}]{COMA_2009A&A...494..403L}
{Lederer}, M.~T. \& {Aringer}, B. 2009, \aap, 494, 403

\bibitem[{{Leinert} {et~al.}(2003){Leinert}, {Graser}, {Przygodda}, {Waters},
  {Perrin}, {Jaffe}, {Lopez}, {Bakker}, {B{\"o}hm}, {Chesneau}, {Cotton},
  {Damstra}, {de Jong}, {Glazenborg-Kluttig}, {Grimm}, {Hanenburg}, {Laun},
  {Lenzen}, {Ligori}, {Mathar}, {Meisner}, {Morel}, {Morr}, {Neumann}, {Pel},
  {Schuller}, {Rohloff}, {Stecklum}, {Storz}, {von der L{\"u}he}, \&
  {Wagner}}]{MIDI}
{Leinert}, C., {Graser}, U., {Przygodda}, F., {et~al.} 2003, \apss, 286, 73

\bibitem[{{Lloyd Evans}(1991)}]{Lloyd_evans_1991MNRAS.248..479L}
{Lloyd Evans}, T. 1991, \mnras, 248, 479

\bibitem[{{Lopez} {et~al.}(2022){Lopez}, {Lagarde, S.}, {Petrov, R. G.},
  {Jaffe, W.}, {Antonelli, P.}, {Allouche, F.}, {Berio, P.}, {Matter, A.},
  {Meilland, A.}, {Millour, F.}, {Robbe-Dubois, S.}, {Henning, Th.}, {Weigelt,
  G.}, {Glindemann, A.}, {Agocs, T.}, {Bailet, Ch.}, {Beckmann, U.},
  {Bettonvil, F.}, {van Boekel, R.}, {Bourget, P.}, {Bresson, Y.}, {Bristow,
  P.}, {Cruzal\`ebes, P.}, {Eldswijk, E.}, {Fante\"{\i} Caujolle, Y.},
  {Gonz\'alez Herrera, J. C.}, {Graser, U.}, {Guajardo, P.}, {Heininger, M.},
  {Hofmann, K.-H.}, {Kroes, G.}, {Laun, W.}, {Lehmitz, M.}, {Leinert, C.},
  {Meisenheimer, K.}, {Morel, S.}, {Neumann, U.}, {Paladini, C.}, {Percheron,
  I.}, {Riquelme, M.}, {Schoeller, M.}, {Stee, Ph.}, {Venema, L.}, {Woillez,
  J.}, {Zins, G.}, {\'Abrah\'am, P.}, {Abadie, S.}, {Abuter, R.}, {Accardo,
  M.}, {Adler, T.}, {Alonso, J.}, {Augereau, J.-C.}, {B\"ohm, A.}, {Bazin, G.},
  {Beltran, J.}, {Bensberg, A.}, {Boland, W.}, {Brast, R.}, {Burtscher, L.},
  {Castillo, R.}, {Chelli, A.}, {Cid, C.}, {Clausse, J.-M.}, {Connot, C.},
  {Conzelmann, R. D.}, {Danchi, W.-C.}, {Delbo, M.}, {Drevon, J.}, {Dominik,
  C.}, {van Duin, A.}, {Ebert, M.}, {Eisenhauer, F.}, {Flament, S.}, {Frahm,
  R.}, {G\'amez Rosas, V.}, {Gabasch, A.}, {Gallenne, A.}, {Garces, E.},
  {Girard, P.}, {Glazenborg, A.}, {Gont\'e, F. Y. J.}, {Guitton, F.}, {de Haan,
  M.}, {Hanenburg, H.}, {Haubois, X.}, {Hocd\'e, V.}, {Hogerheijde, M.}, {ter
  Horst, R.}, {Hron, J.}, {Hummel, C. A.}, {Hubin, N.}, {Huerta, R.}, {Idserda,
  J.}, {Isbell, J. W.}, {Ives, D.}, {Jakob, G.}, {Jask\'o, A.}, {Jochum, L.},
  {Klarmann, L.}, {Klein, R.}, {Kragt, J.}, {Kuindersma, S.}, {Kokoulina, E.},
  {Labadie, L.}, {Lacour, S.}, {Leftley, J.}, {Le Poole, R.}, {Lizon, J.-L.},
  {Lopez, M.}, {Lykou, F.}, {M\'erand, A.}, {Marcotto, A.}, {Mauclert, N.},
  {Maurer, T.}, {Mehrgan, L. H.}, {Meisner, J.}, {Meixner, K.}, {Mellein, M.},
  {Menut, J. L.}, {Mohr, L.}, {Mosoni, L.}, {Navarro, R.}, {Nu\ss{}baum, E.},
  {Pallanca, L.}, {Pantin, E.}, {Pasquini, L.}, {Phan Duc, T.}, {Pott, J.-U.},
  {Pozna, E.}, {Richichi, A.}, {Ridinger, A.}, {Rigal, F.}, {Rivinius, Th.},
  {Roelfsema, R.}, {Rohloff, R.-R.}, {Rousseau, S.}, {Salabert, D.}, {Schertl,
  D.}, {Schuhler, N.}, {Schuil, M.}, {Shabun, K.}, {Soulain, A.}, {Stephan,
  C.}, {Toledo, P.}, {Tristram, K.}, {Tromp, N.}, {Vakili, F.}, {Varga, J.},
  {Vinther, J.}, {Waters, L. B. F. M.}, {Wittkowski, M.}, {Wolf, S.}, {Wrhel,
  F.}, \& {Yoffe, G.}}]{MATISSE}
{Lopez}, B., {Lagarde, S.}, {Petrov, R. G.}, {et~al.} 2022, A\&A, 659, A192

\bibitem[{{Markwardt}(2009)}]{Non_linear_least_square_2009ASPC..411..251M}
{Markwardt}, C.~B. 2009, in Astronomical Society of the Pacific Conference
  Series, Vol. 411, Astronomical Data Analysis Software and Systems XVIII, ed.
  D.~A. {Bohlender}, D.~{Durand}, \& P.~{Dowler}, 251

\bibitem[{{Mathis} {et~al.}(1977){Mathis}, {Rumpl}, \&
  {Nordsieck}}]{MNR_1977ApJ...217..425M}
{Mathis}, J.~S., {Rumpl}, W., \& {Nordsieck}, K.~H. 1977, \apj, 217, 425

\bibitem[{{Millan-Gabet} {et~al.}(2003){Millan-Gabet}, {Pedretti}, {Monnier},
  {Traub}, {Schloerb}, {Carleton}, {Ragland}, {Lacasse}, {Danchi}, {Tuthill},
  {Perrin}, \& {Coud{\'e} du Foresto}}]{IOTA_interf}
{Millan-Gabet}, R., {Pedretti}, E., {Monnier}, J.~D., {et~al.} 2003, in Society
  of Photo-Optical Instrumentation Engineers (SPIE) Conference Series, Vol.
  4838, Interferometry for Optical Astronomy II, ed. W.~A. {Traub}, 202--209

\bibitem[{{Mohamed} \& {Podsiadlowski}(2012)}]{Mohamed2012BaltA..21...88M}
{Mohamed}, S. \& {Podsiadlowski}, P. 2012, Baltic Astronomy, 21, 88

\bibitem[{{Paladini} {et~al.}(2009){Paladini}, {Aringer}, {Hron}, {Nowotny},
  {Sacuto}, \& {H{\"o}fner}}]{Paladini_model_radius_2009A&A...501.1073P}
{Paladini}, C., {Aringer}, B., {Hron}, J., {et~al.} 2009, \aap, 501, 1073

\bibitem[{{Paladini} {et~al.}(2018){Paladini}, {Baron}, {Jorissen}, {Le
  Bouquin}, {Freytag}, {van Eck}, {Wittkowski}, {Hron}, {Chiavassa}, {Berger},
  {Siopis}, {Mayer}, {Sadowski}, {Kravchenko}, {Shetye}, {Kerschbaum},
  {Kluska}, \& {Ramstedt}}]{2018Natur.553..310P}
{Paladini}, C., {Baron}, F., {Jorissen}, A., {et~al.} 2018, \nat, 553, 310

\bibitem[{{Pantin}(2010)}]{Pantin_2010PhDT.......249P}
{Pantin}, E. 2010, PhD thesis, CEA Saclay, Service d'Astrophysique

\bibitem[{{Pegourie}(1988)}]{SiC_1988A&A...194..335P}
{Pegourie}, B. 1988, \aap, 194, 335

\bibitem[{{Pitman} {et~al.}(2008){Pitman}, {Hofmeister}, {Corman}, \&
  {Speck}}]{RefSiC_2008A&A...483..661P}
{Pitman}, K.~M., {Hofmeister}, A.~M., {Corman}, A.~B., \& {Speck}, A.~K. 2008,
  \aap, 483, 661

\bibitem[{{Planquart} {et~al.}(2024){Planquart}, {Jorissen}, {Escorza},
  {Verhamme}, \& {Van Winckel}}]{self_citation}
{Planquart}, L., {Jorissen}, A., {Escorza}, A., {Verhamme}, O., \& {Van
  Winckel}, H. 2024, \aap, 682, A143

\bibitem[{{Ramstedt} {et~al.}(2014){Ramstedt}, {Mohamed}, {Vlemmings},
  {Maercker}, {Montez}, {Baudry}, {De Beck}, {Lindqvist}, {Olofsson},
  {Humphreys}, {Jorissen}, {Kerschbaum}, {Mayer}, {Wittkowski}, {Cox},
  {Lagadec}, {Leal-Ferreira}, {Paladini}, {P{\'e}rez-S{\'a}nchez}, \&
  {Sacuto}}]{Mira_ALMA_2014A&A...570L..14R}
{Ramstedt}, S., {Mohamed}, S., {Vlemmings}, W.~H.~T., {et~al.} 2014, \aap, 570,
  L14

\bibitem[{{Rouleau} \& {Martin}(1991)}]{AmC_1991ApJ...377..526R}
{Rouleau}, F. \& {Martin}, P.~G. 1991, \apj, 377, 526

\bibitem[{{Sacuto} {et~al.}(2011){Sacuto}, {Aringer}, {Hron}, {Nowotny},
  {Paladini}, {Verhoelst}, \& {H{\"o}fner}}]{Sacuto_2011A&A...525A..42S}
{Sacuto}, S., {Aringer}, B., {Hron}, J., {et~al.} 2011, \aap, 525, A42

\bibitem[{{Sahai} {et~al.}(2008){Sahai}, {Findeisen}, {Gil de Paz}, \&
  {S{\'a}nchez Contreras}}]{Sahai_2008ApJ...689.1274S}
{Sahai}, R., {Findeisen}, K., {Gil de Paz}, A., \& {S{\'a}nchez Contreras}, C.
  2008, \apj, 689, 1274

\bibitem[{{Sahai} {et~al.}(2022){Sahai}, {Huang}, {Scibelli}, {Morris},
  {Hinkle}, \& {Lee}}]{SahaiALMA}
{Sahai}, R., {Huang}, P.~S., {Scibelli}, S., {et~al.} 2022, \apj, 929, 59

\bibitem[{{Sahai} {et~al.}(2019){Sahai}, {Rajagopal}, {Hinkle}, {Joyce}, \&
  {Morris}}]{Sahai_2019IAUS..343..495S}
{Sahai}, R., {Rajagopal}, J., {Hinkle}, K., {Joyce}, R., \& {Morris}, M. 2019,
  IAU Symposium, 343, 495

\bibitem[{{Sahai} {et~al.}(2016){Sahai}, {Scibelli}, \& {Morris}}]{Sahai_HST}
{Sahai}, R., {Scibelli}, S., \& {Morris}, M.~R. 2016, \apj, 827, 92

\bibitem[{{Sahai} {et~al.}(2009){Sahai}, {Sugerman}, \&
  {Hinkle}}]{referee_suggestion_Sahai_CO_outflow_2_2009ApJ...699.1015S}
{Sahai}, R., {Sugerman}, B. E.~K., \& {Hinkle}, K. 2009, \apj, 699, 1015

\bibitem[{{Sahai} \&
  {Wannier}(1988)}]{referee_suggestion_Sahai_CO_outflow_paper_1988A&A...201L...9S}
{Sahai}, R. \& {Wannier}, P.~G. 1988, \aap, 201, L9

\bibitem[{{Sargent} {et~al.}(2010){Sargent}, {Srinivasan}, {Meixner}, {Kemper},
  {Tielens}, {Speck}, {Matsuura}, {Bernard}, {Hony}, {Gordon}, {Indebetouw},
  {Marengo}, {Sloan}, \& {Woods}}]{Sargent_2010ApJ...716..878S}
{Sargent}, B.~A., {Srinivasan}, S., {Meixner}, M., {et~al.} 2010, \apj, 716,
  878

\bibitem[{{Schlafly} {et~al.}(2020){Schlafly}, {Meisner}, \&
  {Green}}]{wise_catalog_2020yCat.2363....0S}
{Schlafly}, E.~F., {Meisner}, A.~M., \& {Green}, G.~M. 2020, VizieR Online Data
  Catalog, II/363

\bibitem[{{Thi{\'e}baut}(2008)}]{MiRA}
{Thi{\'e}baut}, E. 2008, in Society of Photo-Optical Instrumentation Engineers
  (SPIE) Conference Series, Vol. 7013, Optical and Infrared Interferometry, ed.
  M.~{Sch{\"o}ller}, W.~C. {Danchi}, \& F.~{Delplancke}, 70131I

\bibitem[{{Townes} {et~al.}(2011){Townes}, {Wishnow}, \& {Ravi}}]{Townes2011}
{Townes}, C.~H., {Wishnow}, E.~H., \& {Ravi}, V. 2011, \pasp, 123, 1370

\bibitem[{{Tsuji} {et~al.}(1988){Tsuji}, {Unno}, {Kaifu}, {Izumiura}, {Ukita},
  {Cho}, \& {Koyama}}]{Tsuji}
{Tsuji}, T., {Unno}, W., {Kaifu}, N., {et~al.} 1988, \apjl, 327, L23

\bibitem[{{Verhamme} {et~al.}(2024){Verhamme}, {Kluska}, {Ferreira}, {Bollen},
  {De Prins}, {Kamath}, \& {Van Winckel}}]{Jet_post_AGB_2024arXiv240116370V}
{Verhamme}, O., {Kluska}, J., {Ferreira}, J., {et~al.} 2024, arXiv e-prints,
  arXiv:2401.16370

\bibitem[{{Vlemmings} {et~al.}(2003){Vlemmings}, {van Langevelde}, {Diamond},
  {Habing}, \& {Schilizzi}}]{WHya_distance_2003A&A...407..213V}
{Vlemmings}, W.~H.~T., {van Langevelde}, H.~J., {Diamond}, P.~J., {Habing},
  H.~J., \& {Schilizzi}, R.~T. 2003, \aap, 407, 213

\bibitem[{{Weigelt} {et~al.}(2016){Weigelt}, {Hofmann}, {Schertl}, {Clementel},
  {Corcoran}, {Damineli}, {de Wit}, {Grellmann}, {Groh}, {Guieu}, {Gull},
  {Heininger}, {Hillier}, {Hummel}, {Kraus}, {Madura}, {Mehner}, {M{\'e}rand},
  {Millour}, {Moffat}, {Ohnaka}, {Patru}, {Petrov}, {Rengaswamy}, {Richardson},
  {Rivinius}, {Sch{\"o}ller}, {Teodoro}, \&
  {Wittkowski}}]{etaCar_AMBER_2016A&A...594A.106W}
{Weigelt}, G., {Hofmann}, K.~H., {Schertl}, D., {et~al.} 2016, \aap, 594, A106

\bibitem[{{Weigelt} {et~al.}(2021){Weigelt}, {Hofmann}, {Schertl}, {Lopez},
  {Petrov}, {Lagarde}, {Berio}, {Jaffe}, {Henning}, {Millour}, {Meilland},
  {Allouche}, {Robbe-Dubois}, {Matter}, {Cruzal{\`e}bes}, {Hillier}, {Russell},
  {Madura}, {Gull}, {Corcoran}, {Damineli}, {Moffat}, {Morris}, {Richardson},
  {Paladini}, {Sch{\"o}ller}, {M{\'e}rand}, {Glindemann}, {Beckmann},
  {Heininger}, {Bettonvil}, {Zins}, {Woillez}, {Bristow}, {Sanchez-Bermudez},
  {Ohnaka}, {Kraus}, {Mehner}, {Wittkowski}, {Hummel}, {Stee}, {Vakili},
  {Hartman}, {Navarete}, {Hamaguchi}, {Espinoza-Galeas}, {Stevens}, {van
  Boekel}, {Wolf}, {Hogerheijde}, {Dominik}, {Augereau}, {Pantin}, {Waters},
  {Meisenheimer}, {Varga}, {Klarmann}, {G{\'a}mez Rosas}, {Burtscher},
  {Leftley}, {Isbell}, {Hocd{\'e}}, {Yoffe}, {Kokoulina}, {Hron}, {Groh},
  {Kreplin}, {Rivinius}, {de Wit}, {Danchi}, {Domiciano de Souza}, {Drevon},
  {Labadie}, {Connot}, {Nu{\ss}baum}, {Lehmitz}, {Antonelli}, {Graser}, \&
  {Leinert}}]{eta_Car_MATISSE_2021A&A...652A.140W}
{Weigelt}, G., {Hofmann}, K.~H., {Schertl}, D., {et~al.} 2021, \aap, 652, A140

\bibitem[{{Wittkowski} {et~al.}(2017){Wittkowski}, {Hofmann}, {H{\"o}fner}, {Le
  Bouquin}, {Nowotny}, {Paladini}, {Young}, {Berger}, {Brunner}, {de
  Gregorio-Monsalvo}, {Eriksson}, {Hron}, {Humphreys}, {Lindqvist}, {Maercker},
  {Mohamed}, {Olofsson}, {Ramstedt}, \&
  {Weigelt}}]{RScl_Wittkowski2017A&A...601A...3W}
{Wittkowski}, M., {Hofmann}, K.~H., {H{\"o}fner}, S., {et~al.} 2017, \aap, 601,
  A3

\bibitem[{{Zhao-Geisler} {et~al.}(2012){Zhao-Geisler}, {Quirrenbach},
  {K{\"o}hler}, \& {Lopez}}]{Zhao_Geisler}
{Zhao-Geisler}, R., {Quirrenbach}, A., {K{\"o}hler}, R., \& {Lopez}, B. 2012,
  \aap, 545, A56

\bibitem[{{Zhao-Geisler} {et~al.}(2011){Zhao-Geisler}, {Quirrenbach},
  {K{\"o}hler}, {Lopez}, \& {Leinert}}]{Zhao_Geisler1}
{Zhao-Geisler}, R., {Quirrenbach}, A., {K{\"o}hler}, R., {Lopez}, B., \&
  {Leinert}, C. 2011, \aap, 530, A120

\end{thebibliography}

\begin{appendix} 
\section{Log of the observations}
\label{appendix:log}
Table \ref{tab:log_observation} summarizes the log of MATISSE observations for V~Hya. Table \ref{tab:cali_info} lists the calibrators used and their characteristics.

\begin{table*}[h]
\centering
\caption{Log of observations used for V~Hya. The column labelled "VLTI config." refers to the AT stations, $\tau_0$ is the coherence time in the visible and FT refers to the fringe tracking ratio, a FITS keyword providing the percentage of fringes tracked by the MATISSE master-band ($L-$band in the specific case) during the observations. The latter is an indicator of the data quality.}
\begin{tabular}{ccllcccc}
\hline \hline
Date & Time [UT]&	VLTI config.&		Object	&Seeing	[\arcsec]&$\tau_0$ [ms]	&FT  \\
\hline

2022-02-25	&04:22:16&	A0-G1-J2-K0&HD 98993&	0.55&	15.33	&1\\
    &	04:44:46&	&	V~Hya&	0.55&	19.49&0.98\\
    &	04:58:09&	&	$\alpha$ Hya&	0.53&	16.38&1\\
2022-02-27	&06:43:12& A0-G1-J2-K0	&HD 98993	&0.41	&11.98&1 \\
    &		07:05:36 &	&V~Hya&	0.35&	12.10& 0.99\\
    &	07:27:56& &	$\epsilon$ Crv&	0.40&	15.08& 1\\
2022-02-28&	05:21:38&	A0-G1-J2-J3&	HD 98993&	0.55&	8.02& 1\\
    &	05:43:24&	&	V~Hya&	0.67&	5.41&1\\
	&	05:56:18&	&	$\alpha$ Hya&	0.61&	6.45&1\\
 
2022-03-02&	07:23:51&	A0-G1-J2-J3&	HD 98993&	0.43&	14.07&1\\
 
	&	07:48:21&	&	V~Hya&	0.58&	12.24&1\\
	&	08:02:42&	&	$\epsilon$ Crv&	0.54&	13.09&1\\

	2022-03-07	&04:47:02 &	K0-G1-D0-J3&HD 98993&	0.43&	10.28&1\\
	&	05:06:53 & &	V~Hya&	0.37&	11.23&0.98\\ 
	&	05:23:09&	 &$\alpha$ Hya&	0.33&	9.56&1\\

	&05:45:10 &	 &HD 98993&	0.28&	10.63&1\\
	&	06:04:02&	 &V~Hya&	0.31&	8.60&1\\
	&	06:18:17&	 &$\alpha$ Hya&	0.30&	13.54&1\\

2022-03-10	&05:05:15&	K0-G2-D0-J3 &HD 98993&	0.77&	5.35&1\\ 
	&	05:16:55&	&	V~Hya&	0.66&	7.23&1\\
	&	05:45:46&	&	$\epsilon$ Crv&	0.45&	12.58&1\\

2022-03-11&	07:02:41&	A0-B2-D0-J3&	HD 98993&	0.45&	7.15&1\\

	&	07:22:25&	&	V~Hya&	0.49&	6.92&0.99\\
	&	07:46:13&	&	$\epsilon$ Crv&	0.38&	7.93&1\\

	2022-03-12&	00:40:55& A0-B2-D0-C1&V~Hya&	1.26&	3.00 &1\\
	&	01:01:58&	&	$\alpha$ Hya&	1.25&  2.32&1 \\

	&	00:53:11&	&	$\alpha$ Hya&	1.26&	2.08&1 \\
	&	01:22:22&	&	V~Hya&	1.11&	2.47&1 \\
	&	01:47:04&	&	$\alpha$ Hya&	1.05&	3.18&1 \\

	&01:37:11&	&	$\alpha$ Hya&	1.02&	3.23&1 \\
	&	02:54:09&	&	V~Hya&    0.85&	3.32& 1\\

	&	03:15:15&	&	$\alpha$ Hya&	0.82&	4.41& 1\\

	&	03:09:03&	&	$\alpha$ Hya&	0.83&	3.32&1 \\
 
&	03:39:05&	&	V~Hya&	0.84&	5.47& 1\\ 
	&	04:00:56&	&	$\alpha$ Hya&	0.83&	5.62& 1\\
	&	04:16:50&	&	V~Hya&	0.93&	4.44& 1\\ 
	&	04:58:25&	&	$\alpha$ Hya&	0.92&	4.47& 1\\

2022-04-12&00:41:38&	A0-G1-J2-K0&	HD 98993&0.88&	2.96&1\\ 
	&	01:01:04&	&	V~Hya&	0.89&	3.53&0.98\\
	&	01:23:31&	&	$\epsilon$ Crv&	0.63&	4.30&1\\

	2022-04-30&	01:52:21&	A0-B2-D0-C1&	$\alpha$ Hya&	1.32&	4.55&1\\
	&	02:11:33&	&	V~Hya&	0.91&	8.83&1\\

	2022-05-30&	00:55:00&	K0-G2-D0-J3&	HD 98993&0.77&	3.33&1\\
	&	01:16:24&	&	V~Hya&	0.66&	3.81&1\\
	&	01:36:40&	&	$\epsilon$ Crv&	0.79&	3.28&1\\
 \hline
\end{tabular}
\label{tab:log_observation}
\end{table*}

\begin{table*}[h]
    \caption{Diameters of the calibrators from II/346 Vizier catalog \citep{JMMC_vizier_II/346}. Flux values from \cite{vizier:II/361}.}
    \centering
    \begin{tabular}{lcrrrr}
    \hline
    \hline
    Name &Spec. Typ. &$\theta_{UD}$ $L$ [mas] & $\theta_{UD}$ $N$ [mas]& $F_L$ [Jy]& $F_N$ [Jy] \\
    \hline
        HD~98993 & K4&2.71 $\pm$ 0.33& 2.73  $\pm$ 0.33& 80 $\pm$ 2& 12$\pm$ 3\\
        HD~81797 ($\alpha$ Hya) & K3&8.71 $\pm$ 0.63& 8.79 $\pm$ 0.63& 948 $\pm$  48&70 $\pm$  8\\
        HD~105707 ($\epsilon$ Crv) & K2 &5.25 $\pm$ 0.42 &5.29 $\pm$ 0.42& 283 $\pm$ 67 & 43$\pm$ 12\\
        \hline
    \end{tabular}
    
    \label{tab:cali_info}
\end{table*}

\section{MATISSE visibilities and closure phases}
\label{appendix:vis2t3}
\begin{figure*}[h]
\centering
    \includegraphics[trim={0 8.5cm 0 0.4cm},clip,width = 0.95\textwidth]{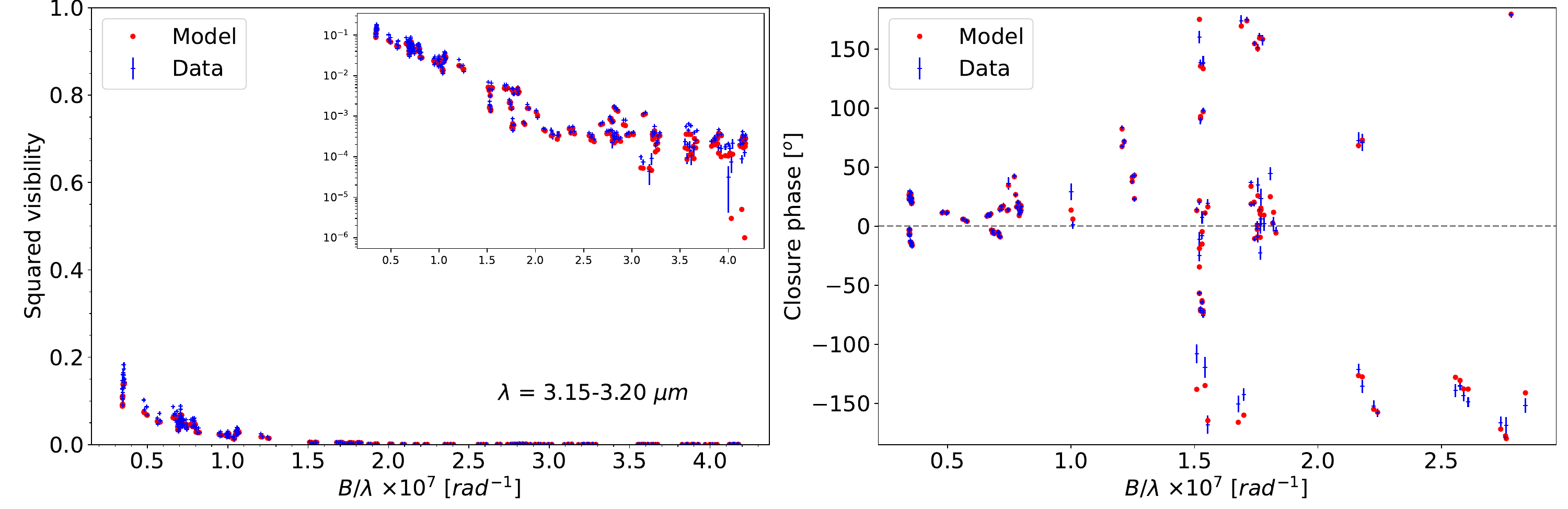}
    \includegraphics[trim={0 8.5cm 0 0.4cm},clip,width = 0.95\textwidth]{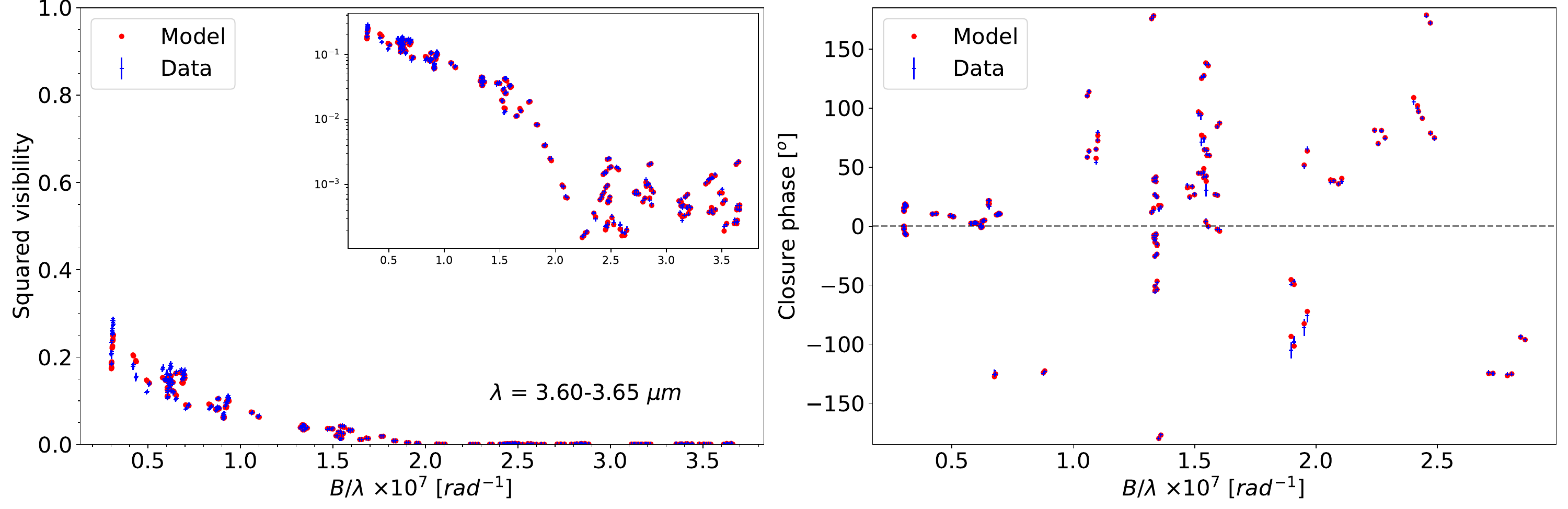}
    \includegraphics[trim={0 8.5cm 0 0.4cm},clip,width = 0.95\textwidth]{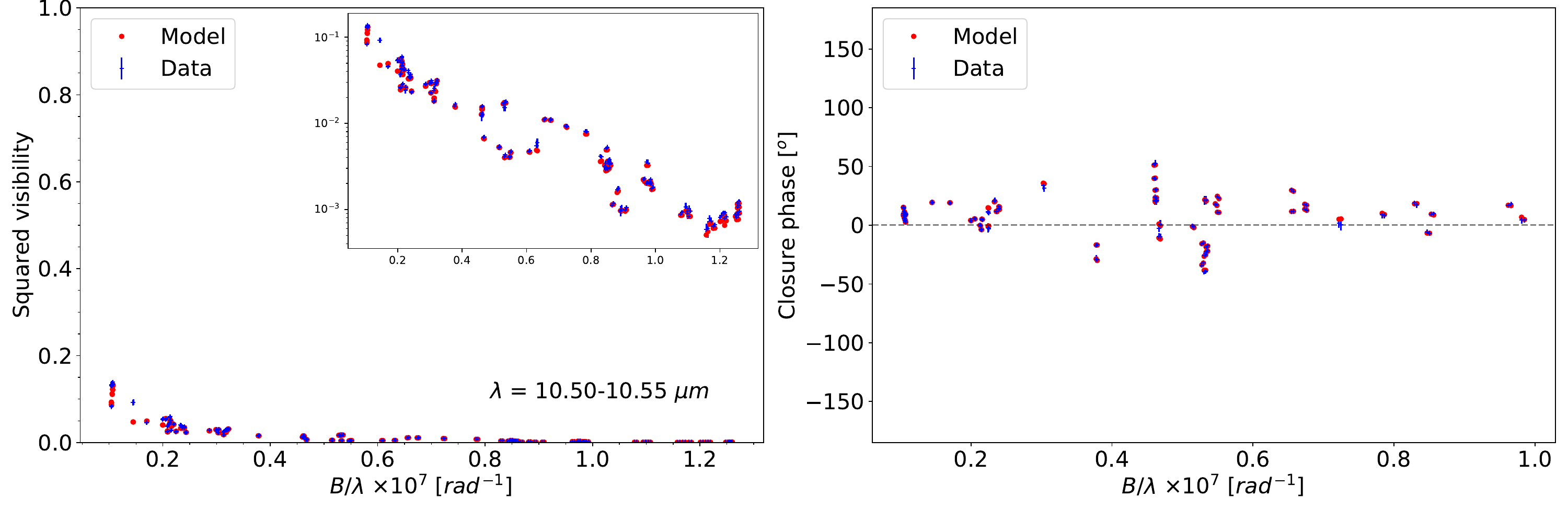}
    \includegraphics[trim={0 8.5cm 0 0.4cm},clip,width = 0.95\textwidth]{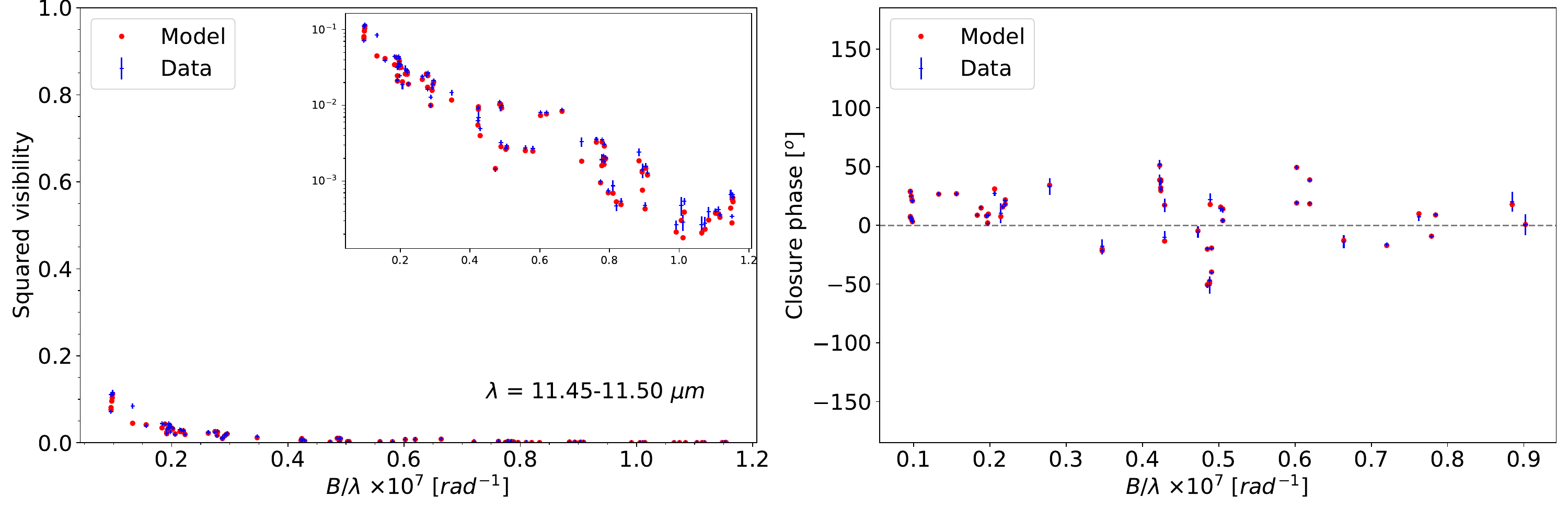}
    \caption{Squared visibilities and closure phases from MATISSE observation (blue error bars). For comparison, the synthetic data from the SQUEEZE image reconstruction is also shown (red dots). The closure phase x-axis represents the longest baseline of each telescope triplet.}
    \label{fig:vis2t3_all}
\end{figure*}

Figure \ref{fig:vis2t3_all} displays the calibrated squared visibilities and closure phases used for the image reconstruction. 

\section{Image reconstruction comparison}
\label{appendix:image_reconstruction}
\begin{figure*}[h]
     \centering
     \includegraphics[width=\textwidth]{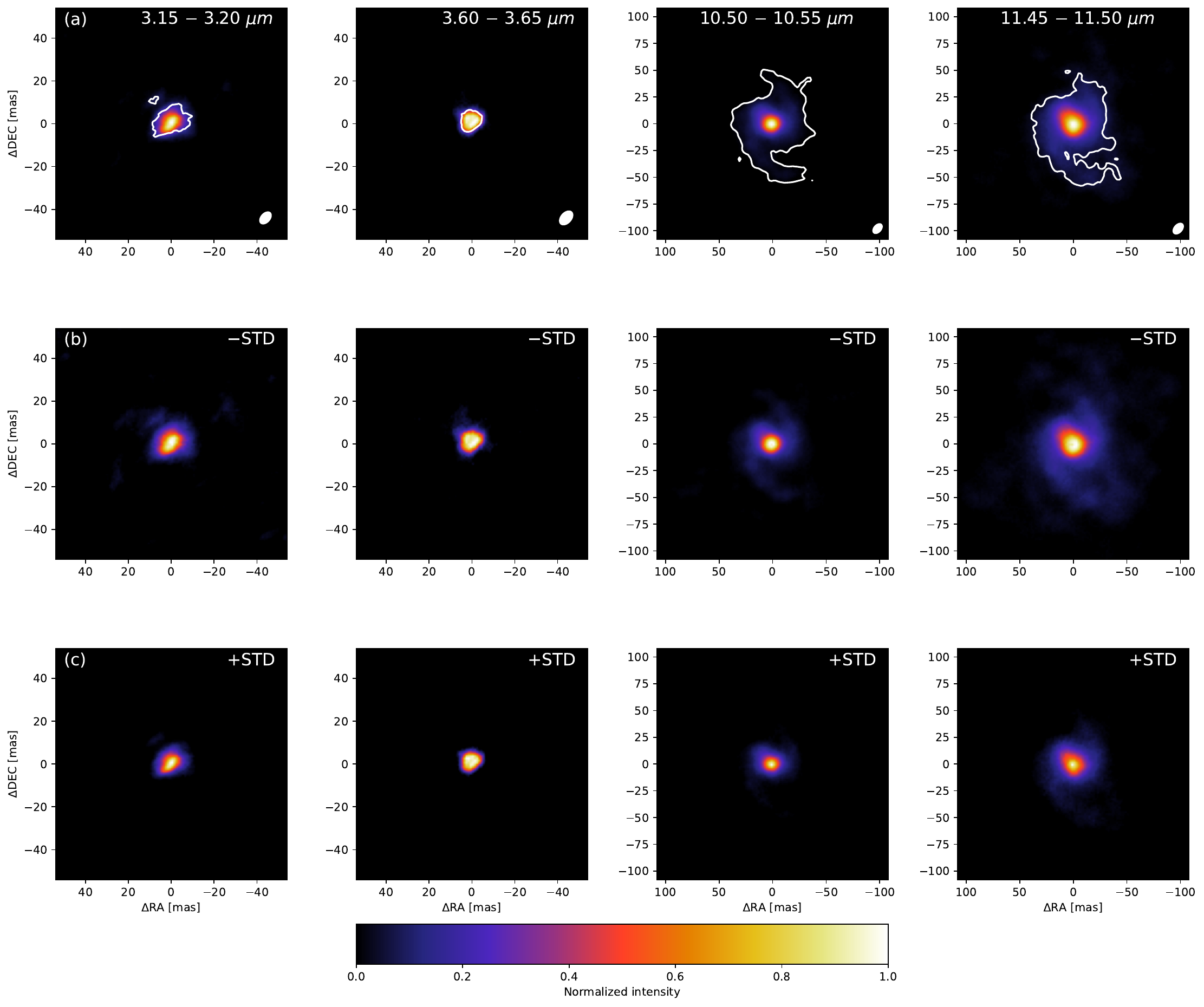}
     \caption{SQUEEZE reconstructed images. (a) Mean image over the chains, (b) Image one standard deviation below the mean image (c) Image one standard deviation above the mean image. The ellipse in the right bottom corner represents the primary synthesised beam. On the upper panels, the white contours are drawn at the 3$\sigma$ level, defined as 3 times the S/N threshold.}
     \label{fig:SQUEEZE_STD}
 \end{figure*}

\begin{figure*}[t]
    \centering
    \includegraphics[width = 0.5\textwidth]{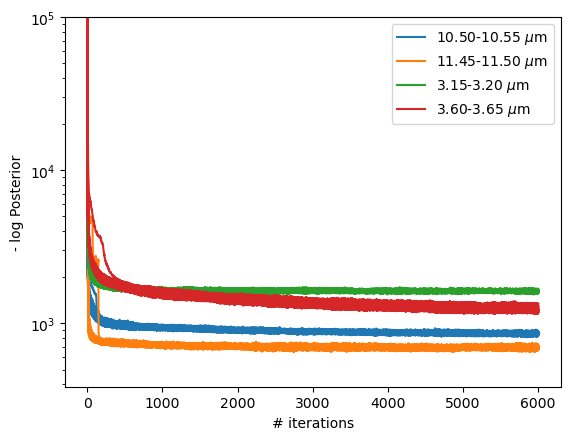}
    \caption{Evolution of the posterior distribution as a function of the iteration for the four SQUEEZE images.}
    \label{fig:evolution_posterior}
\end{figure*}

\begin{figure*}[h]
    \centering
    \includegraphics[width = \textwidth]{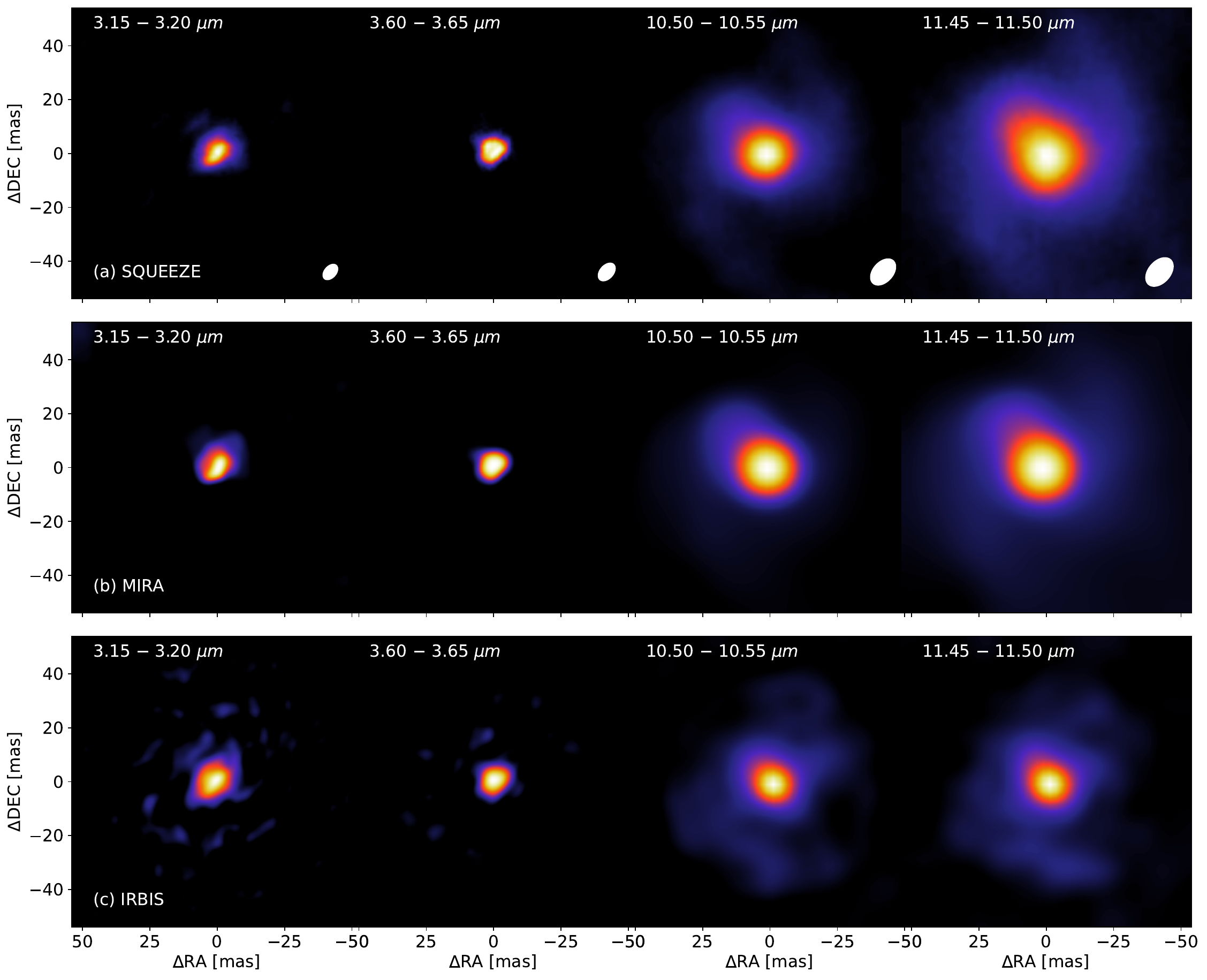}
    \caption{From top to bottom: Image reconstructed with SQUEEZE (top row), with MIRA (middle row) and IRBIS (bottom row) for four different wavelength ranges. The images are convolved by a Gaussian beam of FHWM equal to the angular resolution of the interferometer (from left to right: 2.3~mas, 2.67~mas, 7.75~mas, 8.45~mas).The color scale is the same as in Fig. \ref{fig:SQUEEZE_STD}.}
    \label{fig:compar_plot}
\end{figure*}

\begin{figure*}[h]
    \centering
    \includegraphics[width = 0.8\textwidth]{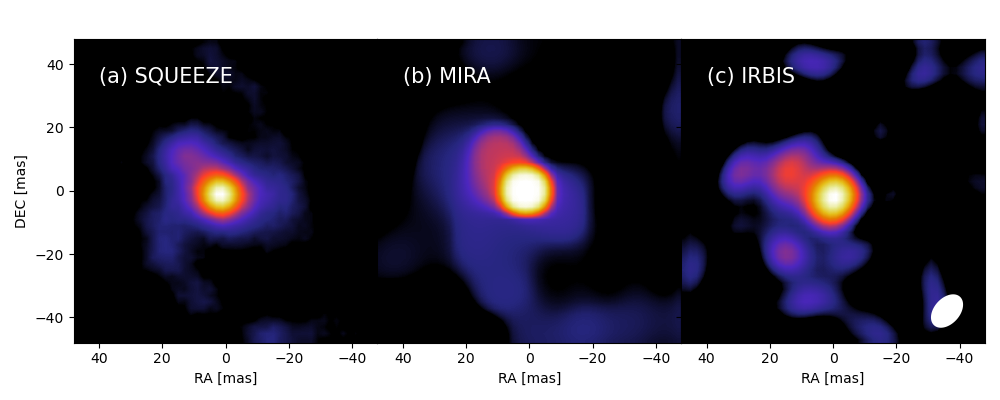}
    \caption{Same as Fig. \ref{fig:compar_plot} but for the image reconstructed in the wavelength range 10.50--11.50~$\mu$m.}
    \label{fig:Mono}
\end{figure*}

Figure \ref{fig:SQUEEZE_STD} displays the mean SQUEEZE images over the chains, together with images obtained at $\pm 1$ standard deviation from the mean. The S/N threshold is defined as the ratio of the mean
image over the standard deviation image. The evolution of the posterior distribution for each chains of the four SQUEEZE images is shown in Fig. \ref{fig:evolution_posterior}. Figures \ref{fig:compar_plot} and \ref{fig:Mono} compare the images obtained with three image reconstruction packages. The SQUEEZE procedure is described in Sect. \ref{sect:image_reconstruction}, the packages MIRA and IRBis and their application are described below. 
To allow direct comparison of the images, the field of view, the pixel resolution and the spectral range are the same for all three methods.

\subsection{MIRA procedure}

MIRA  - Multi-aperture Image Reconstruction Algorithm \citep{MiRA} - follows a gradient-based search to minimize the weighted sum of the likelihood and a regularization term that priors information imposed by the regularizer. The total variation regularization was used with a hyper-parameter, $\mu$, of 10000. The $\mu$ value was set using the L-curve technique for values ranging from 0.1 to 100,000 in a semi-log scale. The initial image is a centered Dirac peak. The final images were obtained after performing bootstrapping.

\subsection{IRBis procedure}
IRBis - image reconstruction software using the bispectrum  \citep{IRBis} - was designed for MATISSE and is included in the mat\_tools pipeline.
It also uses a gradient-based search but includes additional parameters such as two minimisation engines, six regularization functions, a weighting of the ($u,v$)-plane and three cost functions; see \citet{IRBis} for more details. 
All images were obtained using the cost function 1 (equal weighting of the visibilities and the closure phases in the computation of the $\chi^2$) and the reduction engine 2 (L-BFGS-B). Table \ref{tab:IRBIS} summarizes the input parameters used for each image and the $\chi^2$ obtained, for the squared visibility and the closure phase.
\begin{table*}[h]
    \centering
    \caption{IRBis image parameters}
    \begin{tabular}{l|rrrr}
        \hline
    \hline
        $\lambda$ [$\mu$m] & ($u,v$)-plane weight&Reg. Func.&  $\mu$ &[$\chi^2_{V^2}$, $\chi^2_{CP}$] \\
        \hline
        3.15-3.20& 0.5&5& 0.05& [0.9, 1.4]\\
         3.60-3.65& 0.5&5& 0.05&[0.7, 1.6]\\
         10.50-10.55& 0.2&6& 100&[1.7, 2.9]\\
         11.45-11.50& 0.2&2& 100&[1.7, 3.4]\\
         \hline
    \end{tabular}
    
    \label{tab:IRBIS}
\end{table*}

\section{Image simulation}
\label{appendix:image_simulation}
\subsection{Estimate of image noise level}

\begin{figure*}[h]
    \centering
    \includegraphics[width = \textwidth]{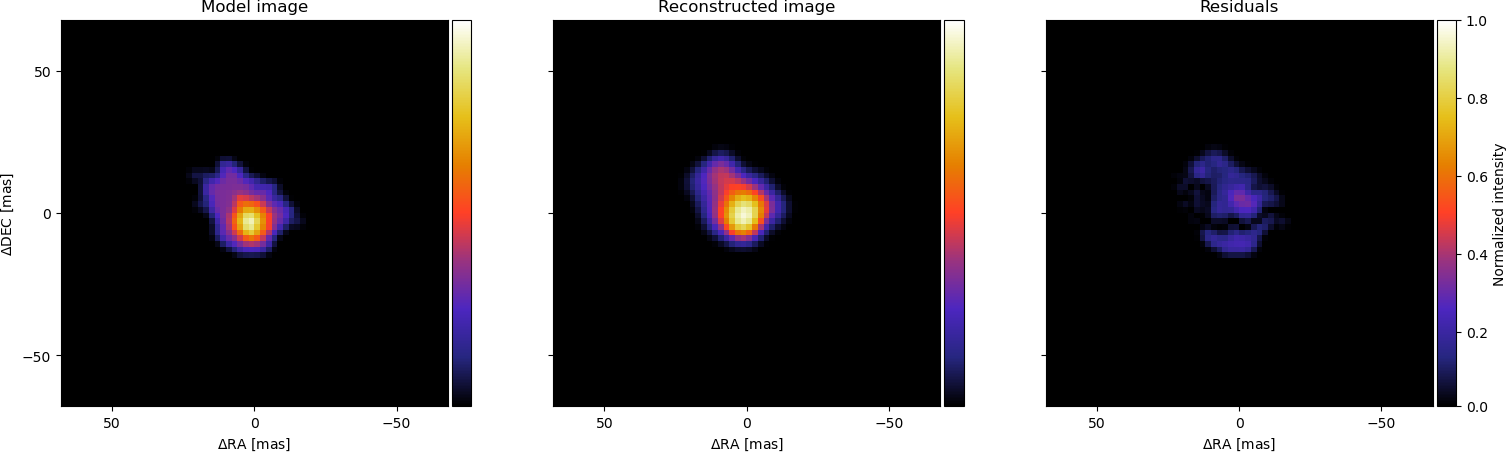}
    \caption{Left: A model image used to assess the reconstruction process. Center: The image reconstructed with the SQUEEZE reconstruction for the 11.45~$\mu$m range. Right: Residual map. The root mean square of the map is 3.06\% of the peak flux level of the original image, this sets the 3$\sigma$ S/N threshold at 9.2 \% on the reconstructed image.}
    \label{fig:Simulation}
\end{figure*}

To quantify the reliability threshold of the images obtained with SQUEEZE, an artificial model was created and an analogous reconstruction was performed. 
The artificial model consists of the median image over the fifty chains obtained for the 11.45~$\mu$m range. The interferometric observables (squared visibilities and closure phases) of the artificial image were created using the same ($u,v$)-coverage \footnote{The observables are extracted using the OIFITS modeler \url{https://amhra.oca.eu/AMHRA/oifits-modeler/input.htm}} and the same uncertainty values as our observations. The artificial image, its reconstruction and the residual map are displayed in Fig. \ref{fig:Simulation}.  

\subsection{Dirty Beam}
\begin{figure*}[h]
    \centering
    \includegraphics[width = 0.9\textwidth]{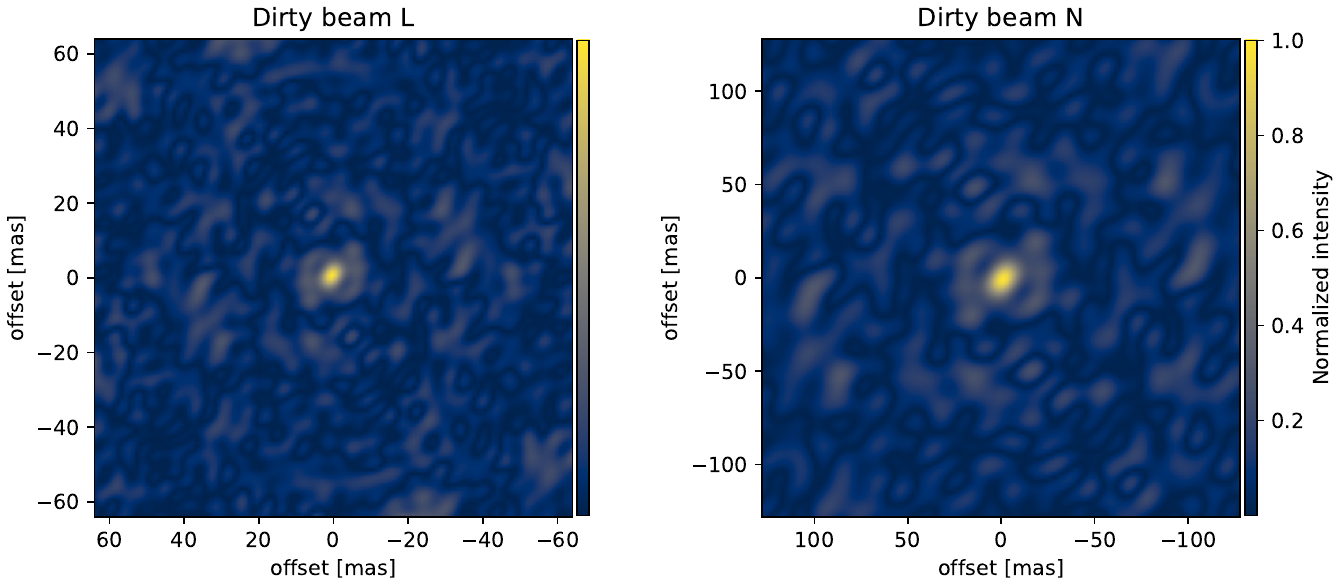}
    \caption{Dirty beam estimated for the MATISSE ($u,v$)-coverage in the $L$- and $N$-bands.}
    \label{fig:DirtyBeam}
\end{figure*}

The reconstruction artefacts are features without any observational counterparts that arise in the inverse problem due to the limited $(u,v)$-plane coverage. 
The best way to identify them is to compute the so-called "dirty beam". The dirty beam is computed by taking the Fourier transform of the ($u,v$)-plane function. The ($u,v$)-plane function was estimated as Dirac peaks at the spatial frequencies of the interferometric measurements. The dirty beam for the $L$- and $N$- band images is shown in Fig. \ref{fig:DirtyBeam}. The shape is the same for both bands but the dimensions of the feature are different as different spatial frequencies are probed. It can be seen that the secondary lobes of the dirty beam have a symmetric spiral shape, with a counter-clockwise winding. 

\subsection{Estimate of image artefacts}
To characterise how the dirty beam impacts the shape of the features obtained, an image error estimate step was performed. First, the interferometric observables of the mirror image were simulated for the two infrared bands. The mirror image is defined as the final reconstructed image (from Fig. \ref{fig:LMN}) but flipped with respect to the vertical axis. The interferometric observables were extracted from the mirror images using the same ($u,v$)-plane coverage and similar values for the uncertainties as for the original observables. The images were reconstructed using SQUEEZE and following the same procedure as described in Sect. \ref{sect:image_reconstruction}: fifty simulated-annealing chains, starting with a different 128$\times$128-pixel random image, were run using the transpectral regularizer for 6000 iterations. The final image was obtained by taking the mean image over the chains. 
Comparing both images allows to identify features arising from the dirty beam. 

\begin{figure*}[h]
    \centering
    \includegraphics[width = 0.75\textwidth]{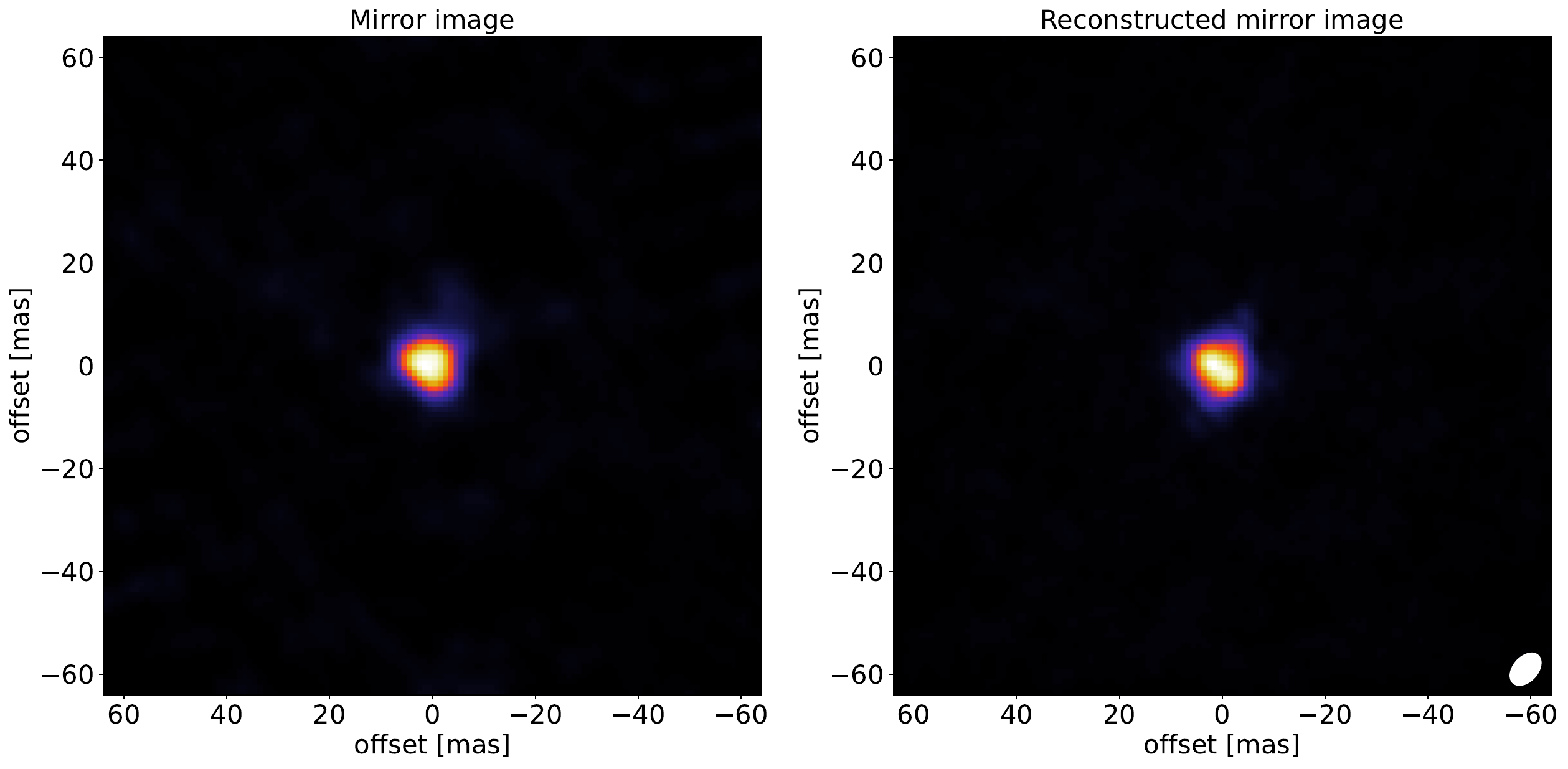}
    \caption{Image simulation in the $L$-band. Left: the simulated mirror image. Right: the image reconstructed with SQUEEZE. North is up, East is left.}
    \label{fig:L_simulation}
\end{figure*}

\begin{figure*}[h]
    \centering
    \includegraphics[width = 0.75\textwidth]{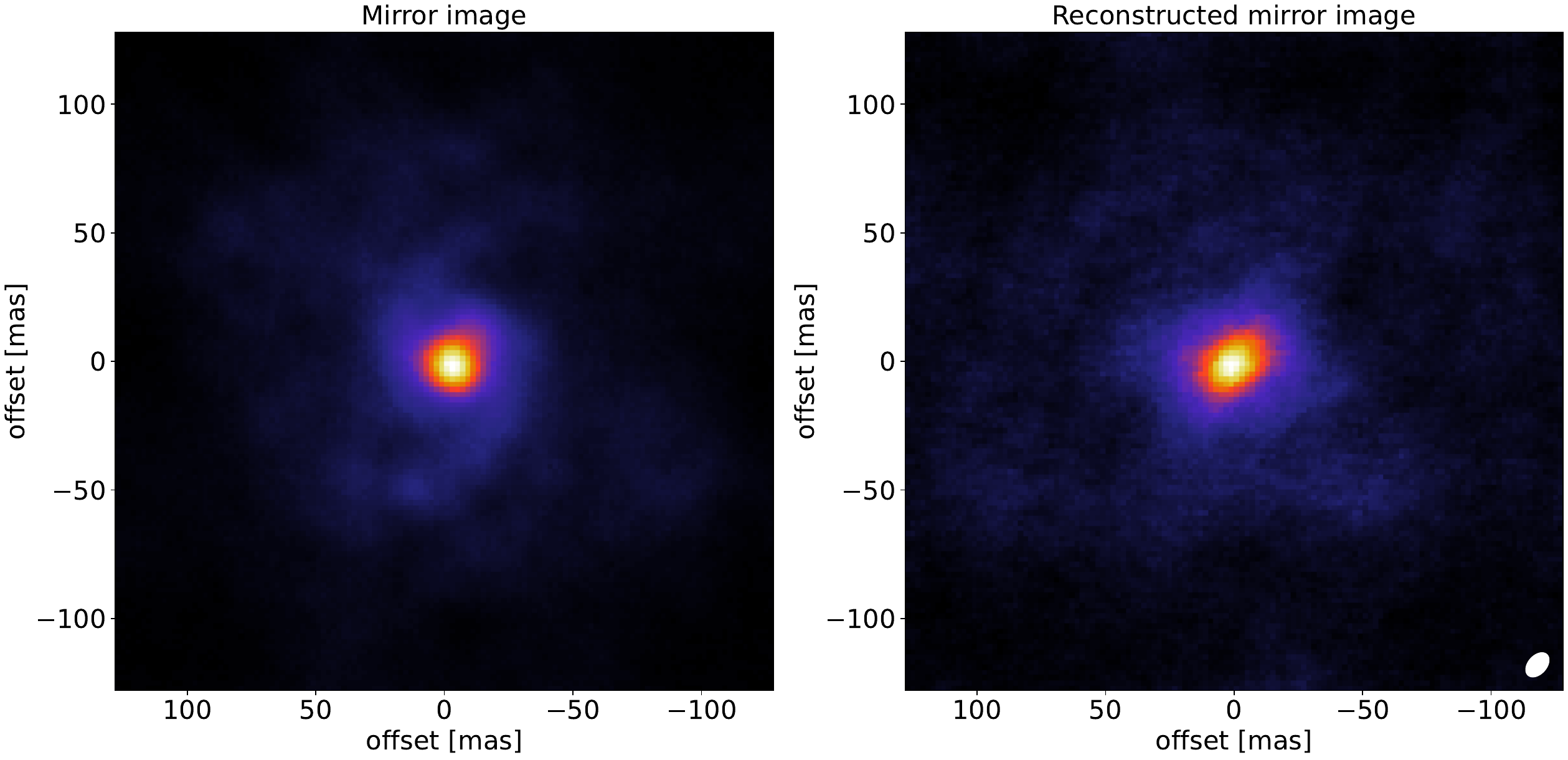}
    \caption{Image simulation in the $N$-band. Left: the simulated mirror image. Right: the image reconstructed with SQUEEZE.  North is up, East is left.}
    \label{fig:N_simulation}
\end{figure*}

The mirror images and their reconstructed counterparts are displayed in Fig. \ref{fig:L_simulation} for the $L$-band and in Fig. \ref{fig:N_simulation} for the $N$-band. 
It can be concluded from Fig. \ref{fig:L_simulation} that the $L$-band mirror image is fully recovered, meaning that the elongated shape is reliable. For Fig. \ref{fig:N_simulation}, the mirror image recovered the expected flipped elongation but the background emission displaying a clockwise spiral in the initial image is not recovered. We can therefore conclude that the shape of the spiral feature is not reliable and arises from the reconstruction process.



\section{Photometric data}
\label{appendix:photometry}

Table \ref{tab:Photometric_data} reports the photometric data (filter name and flux value) used to plot the spectral energy distribution. The data are taken from \url{https://vizier.cfa.harvard.edu/viz-bin/VizieR} with a search radius of 5\arcsec. The WISE photometric fluxes from \cite{wise_catalog_2020yCat.2363....0S} were disregarded due to their  quality factor of 0, implying a heavily saturated detection.
\begin{table*}[h]
    \caption{Photometric data used in the spectral energy distribution of V~Hya.}
    \label{tab:Photometric_data}
    \centering
    \begin{tabular}{lrr}
    \hline
        Filter &  MJD & Flux [Jy] \\
        \hline
        \hline


Tycho:VT & $-$ &2.26 $\pm$0.03\\
Gaia DR2:Gbp& 57204&1.3 $\pm$0.12\\
Gaia DR2:G& 57204&6.84 $\pm$0.32\\
Gaia DR2:Grp& 57204&27.6 $\pm$1.5\\

Gaia DR3:Gbp& 57388&1.7 $\pm$0.07\\
Gaia DR3:G& 57388&13.8 $\pm$0.2\\
Gaia DR3:Grp& 57388&32.0 $\pm$1.9\\

Johnson: V&51242 & 5.77  $\pm$0.16\\
Johnson: J&51242 &353 $\pm$101 \\
Johnson: H& 51242&859 $\pm$161 \\
Johnson: K &51242&1230 $\pm$ 250 \\
2MASS: J&51242&345 $\pm$100 \\
2MASS: H&51242&868 $\pm$162 \\
2MASS: Ks&51242  &1270 $\pm$ 260 \\
IRAS: 12 $\rm \mu m$& $-$&1110 $\pm$60 \\
IRAS: 25 $\mu$m& $-$&460 $\pm$23 \\
IRAS: 60 $\mu$m& $-$&98.9 $\pm$13.8 \\
IRAS: 100 $\mu$m&$-$ &29.9$\pm$ 4.8 \\
AKARI:N60&$-$ &34.3 $\pm$  12.0\\
AKARI:WIDE-S&$-$ &54.0 $\pm$ 16.2\\
AKARI:WIDE-L&$-$ &15.2 $\pm$  9.1\\
AKARI:N160&$-$ &5.1 $\pm$ 3.1\\

    \hline
    \end{tabular}
\end{table*}

\section{VISIR images}
\label{appendix:visirdata}
\begin{figure*}
    \includegraphics[width = 0.308\textwidth]{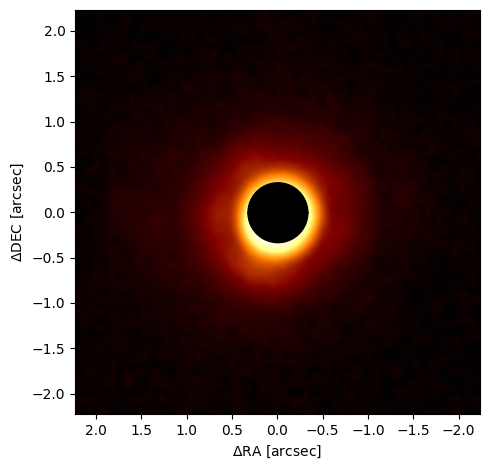}
    \includegraphics[width = 0.308\textwidth]{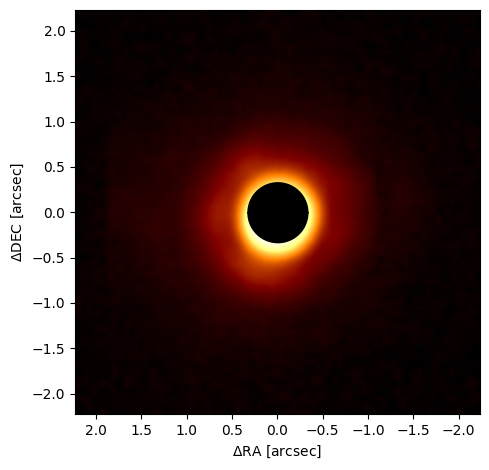}
    \includegraphics[width = 0.384\textwidth]{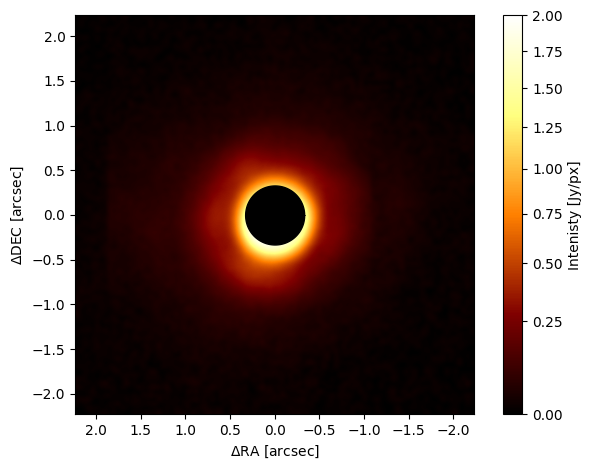}
    \caption{VISIR coronagraphic images for rejection yields of 30\% (left), 60 \% (center) and 90\% (right).}
    \label{fig:visir_rejection}
\end{figure*}

Figure \ref{fig:visir_rejection} shows the final VISIR coronagraphic image, PSF subtracted, for three percentages of retained frames. 

\end{appendix}

\end{document}